\documentclass[fleqn,usenatbib,useAMS]{mnras}

\usepackage{graphicx}	
\usepackage{amsmath}	
\usepackage{multicol}   
\usepackage{bm}		    
\usepackage{pdflscape}	

\usepackage[T1]{fontenc}
\usepackage{ae,aecompl}

\usepackage{newtxtext,newtxmath}
\usepackage{subcaption}

\newcommand{\nH}{n_{\mathrm{H}}}

\title[Non-explosive pre-supernova feedback]{Non-explosive pre-supernova feedback in the COLIBRE model of galaxy formation}

\author[A. Ben\'itez-Llambay et al.]{
Alejandro Ben\'itez-Llambay,$^{1}$\thanks{E-mail: alejandro.benitezllambay@unimib.it (ABL)}
Sylvia Ploeckinger,$^{2}$
Joop Schaye,$^{3}$
Alexander J. Richings,$^{4,5}$
Evgenii Chaikin,$^{3}$
\newauthor
Matthieu Schaller,$^{3,6}$ 
James W. Trayford,$^{7}$
Carlos S. Frenk,$^{8}$
Filip Huško,$^{3}$
and Camila Correa$^{3}$\\
$^{1}$Dipartimento di Fisica G. Occhialini, Universit\`a degli Studi di Milano Bicocca, Piazza della Scienza, 3 I-20126 Milano MI, Italy\\
$^{2}$Department of Astrophysics, University of Vienna, Türkenschanzstrasse 17, 1180 Vienna, Austria\\
$^{3}$Leiden Observatory, Leiden University, PO Box 9513, 2300 RA Leiden, the Netherlands\\
$^{4}$Centre for Data Science, Artificial Intelligence and Modelling, University of Hull, Cottingham Road, Hull, HU6 7RX, UK\\
$^{5}$E. A. Milne Centre for Astrophysics, University of Hull, Cottingham Road, Hull, HU6 7RX, UK\\
$^{6}$Lorentz Institute for Theoretical Physics, Leiden University, PO box 9506, 2300 RA Leiden, the Netherlands\\
$^{7}$Institute of Cosmology and Gravitation, University of Portsmouth, Dennis Sciama Building, Burnaby Road, Portsmouth PO1 3FX, UK\\
$^{8}$Institute for Computational Cosmology, Department of Physics, University of Durham, South Road, Durham, DH1 3LE, UK\\
}

\date{Accepted XXX. Received YYY; in original form ZZZ}

\pubyear{\the\year{}}

\begin{document}
\label{firstpage}
\pagerange{\pageref{firstpage}--\pageref{lastpage}}
\maketitle

\begin{abstract}
We present the implementation and testing of a subgrid non-explosive pre-supernova (NEPS) feedback module for the {\tt COLIBRE} model of galaxy formation. The NEPS module incorporates three key physical processes sourced by young, massive stars that act immediately following star formation: momentum injection from stellar winds and radiation pressure, and thermal energy from photoheating in~\ion{H}{II} regions. The age- and metallicity-dependent energy and momentum budgets are derived from {\tt BPASS} stellar population models and are coupled self-consistently to the local gas properties. We test the model using a suite of smoothed particle hydrodynamics simulations of isolated, unstable gaseous disks at various numerical resolutions (gas particle masses in the range $10^4-10^6$ $\rm M_{\odot}$). We find that the NEPS module successfully regulates star formation by providing pressure support that prevents catastrophic gas collapse. This regulation improves the numerical convergence of star formation rates and disk structure. In our model, feedback from~\ion{H}{II} regions is the dominant regulatory mechanism. Furthermore, we demonstrate a crucial synergy with subsequent supernova feedback; NEPS feedback pre-processes the interstellar medium, creating a more homogeneous environment that moderates the effect of explosive feedback from supernova events. Our NEPS module thus provides a physically motivated and numerically robust framework that mitigates resolution-dependent artefacts and promotes self-regulated galaxy growth. 
\end{abstract}




\section{Introduction}

Within the Lambda Cold Dark Matter ($\Lambda$CDM) cosmological model, galaxies form from gas that cools and sinks into the centers of collapsed dark matter halos \citep{Blumenthal1984}. As the gas collapses and settles into a rotationally supported disk, it fragments and forms the stars that make up a galaxy. Early semi-analytic calculations and cosmological simulations tracking the condensation of baryons into galaxies demonstrated that gas cooling alone produces galaxies that are too massive and compact \citep[e.g.,][]{Navarro1994, Navarro1995}, and also too abundant, particularly at low masses \citep[e.g.,][]{White1991, Cole1991}. These early shortcomings demonstrated compellingly that, for a dissipative galaxy formation model to reproduce observations, additional physical processes were required to prevent excessive cooling and the loss of angular momentum. Obvious candidates included gas heating by supernova (SN) explosions \citep[][and references therein]{White1978, White1991, Cole1994} and the suppression of collisional cooling due to the UV background that keeps the Universe ionized post hydrogen reionization \citep{Efstathiou1992}. The former mechanism alleviates the overcooling and angular momentum loss problems by ejecting and delaying the collapse of disk gas, while the latter (particularly the impact of photoheating), is effective at suppressing star formation in low-mass halos, thereby mitigating the overabundance of faint galaxies \citep[e.g.,][]{Thoul1996,Quinn1996,Bullock2000}.

Soon after these ideas were put forward, cosmological hydrodynamical simulations began incorporating them through subgrid prescriptions for star formation, SN feedback, and photoionization by the UV background. An early example is the simulation of \citet{Katz1996}, which already included most of these elements, paved the way for more advanced models, and highlighted the difficulties inherent to feedback modeling.

It is now well established that SN feedback plays a central role in regulating star formation and in shaping galaxy populations. Not only does it prevent galaxies from becoming too massive, but it also promotes disk formation~\citep[e.g.,][]{Sales2010, Brook2012, Ubler2014}, enables simulations to reproduce the galaxy stellar mass function~\citep[e.g.,][]{Schaye2015}, and establishes key scaling relations such as stellar mass versus size and stellar mass versus halo mass, from which other empirical correlations like the Tully-Fisher relations follow~\citep[e.g.][]{Ludlow2017, Navarro2017}.

Although there is broad agreement on the necessity of feedback, the question of how best to model it has received intense attention over the past two decades. Various strategies have been adopted to address the notorious artificial inefficiency of thermal feedback due to limited numerical resolution: decoupling gas particles from hydrodynamics while imparting kinetic kicks \citep[e.g.,][]{Springel2003}, delaying radiative cooling \citep[e.g.,][]{Stinson2006,Stinson2013}, or injecting large amounts of thermal energy to ensure long cooling times \citep[e.g.,][]{Booth2009, DallaVecchia2012}. Despite their differences, these approaches can all reproduce realistic galaxy populations, but only if carefully calibrated to match observed benchmarks \citep[see, e.g.,][for a recent review]{Crain2023}.

Because subgrid prescriptions necessarily target unresolved physics, they are often sensitive to resolution and may require recalibration when applied to different numerical setups. This limitation is evident in the {\tt EAGLE} model \citep{Schaye2015}, whose parameters were tuned at intermediate resolution. When applied without modification to higher-resolution runs, the model yields galaxies that deviate more strongly from observations~\citep[e.g.,][]{Crain2015}. The problem stems partly from the fact that higher resolution resolves physical processes that low-resolution models must approximate---particularly the collapse of denser (molecular) gas clouds where stars form.

At low resolution, unresolved collapse and the imposition of an effective interstellar medium (ISM) equation of state~\citep[e.g.][]{Springel2003, Schaye2008} force star formation to occur in a relatively smooth medium. In contrast, at high resolution or when the effective equation of state is dropped, stars emerge in dense, clustered, self-gravitating clouds. This transition alters galaxy properties profoundly. Star formation may proceed too rapidly within individual clouds due to short freefall times, while SN feedback can become ineffective if explosions occur in dense gas that cools rapidly, dissipating the injected energy. Moreover, if cloud collapse times are shorter than the lifetimes of massive stars, SNe may explode too late to regulate star formation occurring deep in the densest regions. These challenges make it difficult to achieve numerical convergence without careful (and expensive) recalibration.

This situation echoes, to some extent, the early recognition that SN feedback was required to mitigate the overcooling problem, and highlights the need for additional feedback processes, particularly those operating before the first SNe. This need is especially acute in simulations that resolve the multiphase ISM directly and, therefore, dispense with the effective equation of state used in previous models to impose a numerical pressure floor~\citep[e.g.][]{Hopkins2018, Schaye2025}. While this increased resolution marks a significant advance, it also exacerbates runaway gravitational collapse in the now-resolved cold, dense gas.

Observations and theoretical models highlight the importance of at least three such mechanisms: radiation pressure, stellar winds, and photoionization from young massive stars which together carve out \ion{H}{II} regions and inject momentum into their surroundings. These are collectively termed {\it early} or {\it pre-SN stellar feedback}.

High-resolution simulations have shown the impact of these processes. For instance, \citet{Stinson2013} demonstrated that adding pre-SN thermal pressure to the ISM lowers gas densities, enhances the efficiency of subsequent SN feedback, and reduces galaxy stellar masses. Although implemented via a simplified prescription, their model effectively
mimicked injecting more SN energy than physically reasonable, further regulating star formation. Similarly, \citet{Agertz2013} found that early feedback was essential for reproducing realistic galaxy disks, while \citet{Rosdahl2015} used radiation–hydrodynamic simulations to show that radiation pressure prevents the collapse of dense star-forming clouds, even if its overall effect is weaker than that of SNe. Further simulations by~\cite{Sales2014} have shown that while direct radiation pressure is effective, the feedback from photoionization is substantially more dominant, rapidly heating and expanding gas in a way that further diminishes the impact of radiation pressure.

These studies underline that pre-SN feedback can preprocess the star-forming ISM, stabilizing it against runaway collapse and enabling SN explosions to couple more effectively with the gas. By enhancing the pressure support and dispersing dense clouds, radiation pressure, stellar winds, and photoheating together anticipate the regulatory power of feedback into the pre-supernova phase. Indeed, \citet{Murray2010} quantified the momentum and energy budget of these processes and showed that momentum injection during the pre-SN phase should be able to disperse molecular clouds and limit star formation, thus mitigating numerical overcooling and likely fostering numerical convergence with resolution, consistent with the theoretical framework of radiation-pressure-supported starbursts of~\citet{Thompson2005}

As was the case for SN feedback in the early 2000s, more recent work has begun integrating pre-SN prescriptions into galaxy formation models \citep[e.g.,][and references therein]{Hopkins2014, Nunez2017, Hopkins2018, Shimizu2019, Marinacci2019, Keller2022}. While these models are not yet routinely deployed in large-volume hydrodynamical cosmological simulations, idealized galaxy and disk-patch experiments indicate that they can suppress runaway star formation in dense environments and help establish more realistic star formation efficiencies \citep[e.g.,][]{Kannan2020}. Observational evidence also supports this picture: nearby galaxies reveal that early feedback processes, such as photoionization and stellar winds, are capable of dispersing molecular clouds on timescales comparable to their dynamical times~\citep{Kruijssen2019, Chevance2022}.

Despite these advances, open questions remain. For example, whether radiation pressure and stellar winds inject enough momentum to halt the collapse of giant molecular clouds before the first SNe explode is not fully settled. Moreover, the effectiveness of these processes depends sensitively on resolution: simulations that fail to resolve SN remnants or ISM turbulence may misrepresent their relative impact, making pre-SN feedback less important overall~\citep[e.g.,][]{Smith2019}.

Given the striking evidence of pre-SN processes operating in galaxies, and the ever-growing adoption of these processes in higher-resolution simulations, here we present the implementation of a new subgrid prescription for {\it non-explosive pre-supernova} (NEPS) feedback designed for the {\tt COLIBRE} model of galaxy formation~\citep{Schaye2025, Chaikin2025}. Our model is timely and necessary for {\tt COLIBRE} given its relatively high effective resolution for large volume simulations, which, among several other improvements detailed below, allows the gas to cool down to $10$ K through careful tracking of the thermal balance of the ISM, enabling simulations to resolve the formation of molecular clouds. Our work thus contributes to the ongoing community effort with a model designed for application in large-volume cosmological simulations where resolving the detailed structure of individual star-forming clouds remains computationally prohibitive.

The NEPS feedback model presented here explicitly accounts for momentum and energy inputs from massive stars prior to the onset of core-collapse SNe and is designed to complement the SN feedback channel during the early stages of cloud collapse. We refer the reader to~\cite{Schaye2025} for an application of our NEPS module in cosmological simulations, and focus here on describing its implementation, physics rationale, and its impact on galaxy-scale star formation.

We begin in Sec.~\ref{Sec:NEPS_model} by presenting the physical rationale and numerical implementation of the NEPS model. Sec.~\ref{Sec:Code} briefly describes our numerical code and the {\tt COLIBRE} model of galaxy formation. We then detail the numerical setup used to test the NEPS model in Sec.~\ref{Sec:Simulations}. The results of our simulations are presented in Sec.~\ref{Sec:Results}, followed by a discussion of their implications in Sec.~\ref{Sec:Discussions}. Our conclusions are summarized in Sec.~\ref{Sec:Conclusions}.

\section{The COLIBRE early feedback model}
\label{Sec:NEPS_model}
\begin{figure}
\includegraphics[width=\columnwidth]{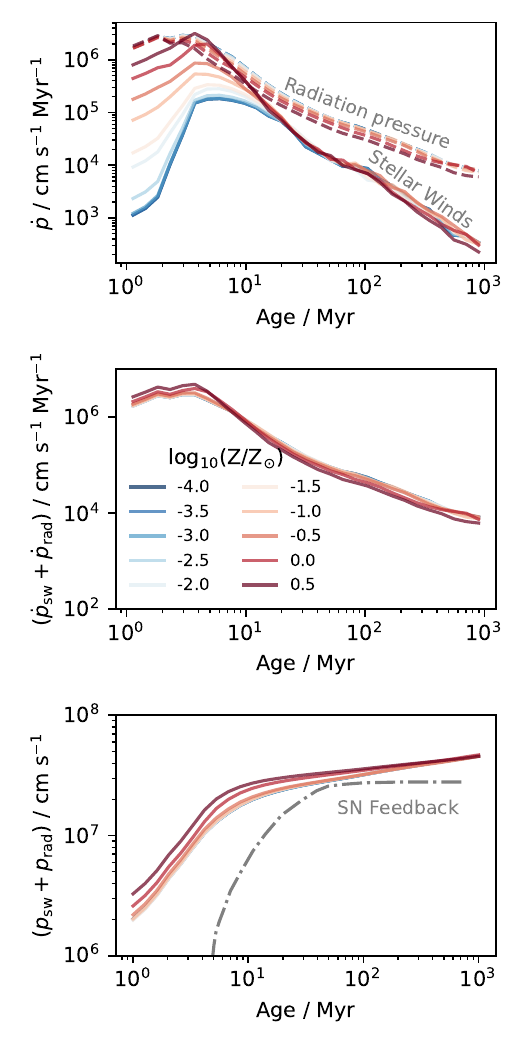}
    \caption{\textit{Top:} Decomposition of the momentum rate, predicted by {\tt BPASS} for a~\protect\cite{Chabrier2003} IMF, into contributions from stellar winds (solid lines) and radiation pressure (dashed lines). While radiation pressure dominates the total momentum budget, only a fraction of this momentum couples effectively to the gas (see Fig.~\ref{fig:fabs_BPASS}). \textit{Middle:} Total specific momentum rate as a function of stellar age (x-axis) and metallicity (coloured lines, as indicated by the legend). \textit{Bottom:} Cumulative total specific momentum as a function of age and metallicity. For comparison, the dashed line indicates the cumulative momentum rate expected from core-collapse SN in {\tt COLIBRE}.}
\label{fig:momentum_stellar_winds}
\end{figure}

The NEPS feedback model is implemented in the {\tt SWIFT} Smoothed Particle Hydrodynamics (SPH) code~\citep{Schaller2024} and is a core component of the {\tt COLIBRE} model of galaxy formation~\citep{Schaye2025}. The physics module described here is designed to be fully consistent with the {\tt HYBRID-CHIMES} module for radiative cooling and chemistry~\citep{Richings2014a,Richings2014b,Ploeckinger2025}. Our NEPS module is designed to capture the effects of three distinct feedback channels that have been largely neglected in previous generations of large-volume hydrodynamical cosmological simulations of galaxy formation. These channels are all sourced by young, short-lived, massive stars: (i) stellar winds, (ii) radiation-pressure-driven winds, and (iii) local photoionization and photoheating. As discussed below, these processes operate soon after stellar populations are born. Although they are applied over a long timescale, most of their energy budget is distributed during the first $\sim 10$ Myr, a timescale that matches the lifetime of a massive O star. As such, all of these processes begin to inject energy into the medium on timescales shorter than the onset of core-collapse SNe. 

We begin by describing the ingredients of the subgrid prescription for our NEPS feedback module.

\subsection{Stellar winds and radiation pressure from young stars}

The intense radiation emitted by young, massive, and short-lived stars drives powerful outflows that carry substantial momentum. In addition, the energetic photons released by these stars can be absorbed by surrounding gas and dust, imparting their momentum and dispersing star-forming gas. We refer to these two channels as stellar wind-driven ($\dot p_{\rm sw}$) and radiation pressure-driven ($\dot p_{\rm rad}$) momentum outflows. We define the total specific momentum rate due to these processes as:
\begin{equation}
    \dot p_{*} (t_{*}, Z_{*}) = \dot p_{\rm sw}(t_*, Z_*) + \dot p_{\rm rad} (t_*, Z_*),
\end{equation}
where we emphasize that the specfic momentum rate, $\dot p_{*}$---i.e., the momentum injected per unit initial stellar mass per unit time---depends on the age of the stellar population, $t_{*}$, and its metallicity, $Z_{*}$.

To compute $\dot p_{\rm sw}$ for a given simple stellar population (SSP), we adopt the {\tt BPASS} stellar population models~\citep{Eldridge2017, Stanway2018}, specifically using the model suite that includes the effects of binary interactions, for the~\cite{Chabrier2003} stellar initial mass function with mass limits between $0.1$-$100 \rm \  M_{\odot}$. {\tt BPASS} provides the stellar wind-driven mass-loss rate, $\dot m_{\rm sw}$, and the associated energy output, $\dot E_{\rm sw}$, as a function of the SSP age and metallicity. Metallicities outside the tabulated range are clipped.

The specific momentum injection rate of the SSP can be expressed in terms of these two quantities as:
\begin{equation}
    \dot p_{\rm sw} (t_*, Z_*)= \displaystyle\frac{1}{m_{*}} \left [ 2 \dot E_{\rm sw} (t_*, Z_*) \dot m_{\rm sw} (t_*, Z_*) \right ]^{1/2}, 
\end{equation}
where $\dot E_{\rm sw}$ and $\dot m_{\rm sw}$ are the total rates for an SSP of mass $m_{*}$.

The momentum injection rate due to radiation pressure is given by:
\begin{equation}
    \dot p_{\rm rad} (t_*, Z_*)= \displaystyle\frac{1}{c} \displaystyle\int_{0}^{\infty} F(\lambda; t_*, Z_*) f_{\rm abs}(\lambda) d\lambda,
\end{equation}
where $c$ is the speed of light, $F(\lambda; t_*, Z_*)$ is the spectral energy distribution per unit stellar mass of the SSP (taken from {\tt BPASS}), which depends on the age and metallicity of the stars and wavelength, $\lambda$, and $f_{\rm abs}(\lambda)$ is the fraction of photons absorbed by the surrounding gas. This absorption fraction depends on gas composition and can be expressed as:
\begin{equation}
    f_{\rm abs}(\lambda) = 1 - e^{\tau(\lambda)},
\end{equation}
where the optical depth, $\tau(\lambda)$, is defined by:
\begin{equation}
    \tau(\lambda) = \displaystyle\sum_{j} \tau_{j}(\lambda) =  \displaystyle\sum_{j} \displaystyle\int_{0}^{L} n_{j}(l) \sigma_j (\lambda) \ dl.
\end{equation}
Here, $n_{j}$ and $\sigma_{j}$ denote the number density and absorption cross-section of the $j$-th species, respectively, and $L$ is the path length through the intervening material, which, for our model, and for consistency with the {\tt COLIBRE} model, is the maximum between the local thermal and turbulent Jeans lengths~\citep[see][]{Ploeckinger2025}. 

Neglecting the detailed density structure of the gas surrounding newly formed stars, we approximate the optical depth as:
\begin{equation}
    \tau_{j} (\lambda) = n_{j} \displaystyle\int_{0}^{L} \sigma_j (\lambda) \ dl =  n_{j} L \sigma_j (\lambda) \equiv N_{j} \sigma_j (\lambda),
    \label{Eq:optical_depth}
\end{equation}
where $N_{j}$ is the column density of the $j$-th species along the path length $L$. For this work, we approximate $n_{j}$ using the local SPH number density.

We note that our model assumes the single-scattering limit, where a photon's momentum is transferred only once upon absorption. In dense, dusty environments, however, a more complex process can occur: UV photons are absorbed and reprocessed into infrared radiation. If the medium is also optically thick to these infrared photons, then they can scatter multiple times before escaping, boosting the total momentum transfer by a factor related to the infrared optical depth. Since our current model neglects this multiple-scattering effect, it provides a conservative estimate and likely underestimates the total radiation momentum coupled to the gas in the most opaque regions. While some studies in low gas surface density environments have found this effect to be modest~\citep[e.g.][]{Rosdahl2015, Kannan2020, Menon2022}, its importance is expected to increase in the very high-density regimes~\citep[e.g.][]{Menon2023}. We therefore caution that our model likely underestimates the full impact of radiation pressure in the most vigorous star-forming regions, and view the inclusion of multi-scattering effects as an avenue for future model development. While not examined here, our wavelength-dependent treatment of radiation pressure establishes a framework for future implementations to capture multi-scattering effects---for example, by enhancing the momentum contribution in the infrared.

For consistency with the hybrid cooling model used in {\tt COLIBRE}, we adopt the~\cite{Ploeckinger2025} estimates for the column densities of neutral and molecular hydrogen, as well as neutral and ionized helium, and dust. These depend on gas density, temperature, metallicity, and redshift, which accounts for the evolution of the background radiation field, and they are used to self-consistently calculate the ionization balance, molecule fractions, and cooling rates. With these column densities we estimate the optical depths using Equation~\ref{Eq:optical_depth}, which allow us to estimate, in turn, the fraction of photons absorbed by the gas. 

We tabulate the specific momentum input from stellar winds and radiation pressure predicted by {\tt BPASS}, as a function of stellar age and metallicity. Using the procedure described above, we then compute the fraction of this momentum absorbed by the gas on a grid spanning density, temperature, metallicity, and time. As explained above, for this calculation we adopt the equilibrium abundances and shielding lengths precomputed for the {\tt COLIBRE} model, as reported in~\cite{Ploeckinger2025}. 

The approximate optical depths for hydrogen and helium are given by:
\begin{align}
    \tau_{\mathrm{H}} (\lambda)& = \sigma_{\mathrm{HI}}(\lambda)
    \begin{cases}
        (N_{\mathrm{HI}} + 2 N_{\mathrm{H_2}})  & \mathrm{if}\, \lambda < \lambda_{i,\mathrm{H2}} \\
        N_{\mathrm{HI}} & \mathrm{if}\,  \lambda \ge \lambda_{i,\mathrm{H2}} \\
    \end{cases}\\
    \tau_{\mathrm{He}}  (\lambda)& = \sigma_{\mathrm{HeI}}(\lambda)
    \begin{cases}
        (N_{\mathrm{HeI}} + 0.75 N_{\mathrm{HeII}})  & \mathrm{if}\, \lambda < \lambda_{i,\mathrm{HeII}} \\
        N_{\mathrm{HeI}} & \mathrm{if}\,  \lambda \ge \lambda_{i,\mathrm{HeII}}.
    \end{cases}
\end{align}
Here, $\sigma_{\ion{H}{I}}$ and $\sigma_{\ion{He}{I}}$ are the neutral hydrogen and helium absorption cross sections, respectively; $N_{\rm HI}$, $N_{\rm H_2}$, $N_{\rm HeI}$, and $N_{\rm HeII}$ are the column densities of neutral and molecular hydrogen and neutral and singly ionized helium, taken from~\cite{Ploeckinger2025}. In practice, we use the maximum of the thermal and turbulent pressure (1D velocity dispersion of $6$ km s$^{-1}$) for the Jeans column density, and use this value to calculate the column densities for ~\ion{H}{I},~\ion{He}{I}, and~\ion{He}{II} with {\tt CHIMES}~\citep{Richings2014a, Richings2014b}, assuming chemistry and ionization equilibrium. 

\begin{figure}
\includegraphics[width=\columnwidth]{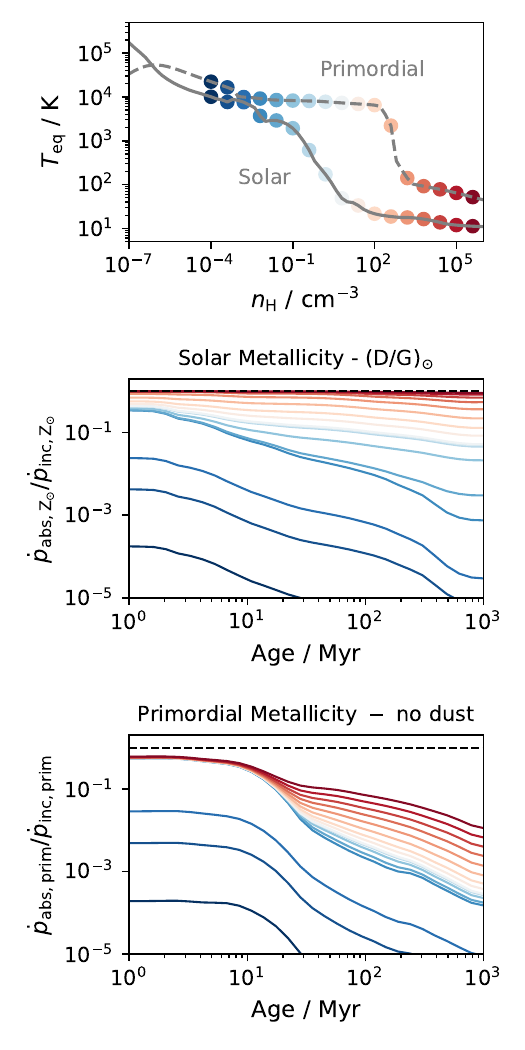}
    \caption{\textit{Top:} Equilibrium temperature as a function of density for solar (solid) and primordial (dashed) metallicity gas, computed from the redshift $z=0$ {\tt COLIBRE} cooling curves. Coloured dots mark the density values illustrated in the lower panels. \textit{Middle:} Ratio of absorbed to incident radiation momentum rate as a function of stellar age (x-axis) for solar metallicity gas, with different colours corresponding to the densities marked above. \textit{Bottom:} Same as the middle panel, but for gas of primordial metallicity. The horizontal dashed lines indicate unity. We adopt a solar dust-to-gas (D/G)$_{\odot}$ ratio and no dust for gas of solar and primordial metallicity, respectively.}
\label{fig:fabs_BPASS}
\end{figure}

For dust absorption, we adopt the~\cite{Draine2003} absorption cross section, $\sigma_{\rm dust} (\lambda)$, for the~\cite{Weingartner2001} carbonaceous-silicate grain model with $R_{V} = A_{V}/(A_{B}-A_{V}) = 3.1$, where $A_{V}$ is the total absorption in the $V$ band and $(A_{B}-A_{V})$ is the so-called selective absorption. The grain abundances have been renormalized by a factor of 0.93 to better match the local ISM, following~\citet{Draine2003}. The dust mass per hydrogen nucleon is $m_{\rm dust/H} = 1.398 \times 10^{-26} \rm g \ H^{-1}$, yielding a dust-to-gas ratio $(D/G)_{\odot} = 5.9 \times 10^{-3}$ for the solar hydrogen mass fraction, $X_{\rm H} = 0.74$. The corresponding dust optical depth, which depends on the local gas properties, is then:
\begin{equation}
    \tau_{\mathrm{dust}}(\lambda) = N_{\mathrm{H}} m_{\mathrm{dust/H}} \frac{D/G}{(D/G)_{\odot}}\kappa_{\mathrm{abs}}(\lambda),
\end{equation}
where $N_{\rm H}$ and  $D/G$ are again taken from~\cite{Ploeckinger2025}.

For the Lyman-Werner band (911.65~\AA\ $< \lambda <$ 1107~\AA), which governs H$_{2}$ self-shielding, we adopt the fitting function for the radiation suppression factor $S_{\mathrm{self}}^{\rm H_{2}}$ from {\tt CHIMES} (equation 3.12 in \citealp{Richings2014a}). This depends on the gas temperature, $T$, the H$_{2}$ column density, $N_{\rm H_{2}}$, and the turbulent velocity dispersion, $\sigma_{\mathrm{turb, 1D}}$, which is assumed constant at $\sigma_{\mathrm{turb,1D}}= 6\,\mathrm{km\,s}^{-1}$ following~\cite{Ploeckinger2025}. The corresponding optical depth is:
\begin{equation}
    \tau_{\rm H_{2}} (\lambda) = - \mathrm{ln} \ S_{\mathrm{self}}^{\rm H_{2}} (\lambda).
\end{equation}

The specific momentum rate that a given SSP can inject through stellar winds and radiation pressure, as a function of time and metallicity, is presented in Fig.~\ref{fig:momentum_stellar_winds}. The top and middle panels show the contributions from radiation pressure and stellar winds and their sum, respectively, whereas the bottom panel shows the resulting total cumulative specific momentum. The momentum is shown in all panels as a function of time, with different colours corresponding to different metallicities. 

Fig.~\ref{fig:momentum_stellar_winds} shows that radiation pressure dominates the total specific momentum rate budget, with stellar winds contributing significantly only at high metallicities and during the first few Myr after formation. For young stellar populations, the total specific momentum available from SSPs reaches up to $\sim 5 \times 10^{7} \ \rm cm \ s^{-1}$ over the sampled timescale, with only a weak dependence on metallicity (bottom panel). Importantly, more than half of this momentum is delivered within the first 10 Myr after birth---a timescale comparable to the lifetimes of the most massive O stars and preceding the peak of the SN rate from the SSP.

Despite the fact that radiation pressure dominates the overall momentum budget, only a fraction of this momentum is actually transferred to the surrounding medium. As discussed above, momentum coupling requires that the gas first absorbs radiation, a condition that is only met in sufficiently dense and neutral gas. 

The two bottom panels of Fig.~\ref{fig:fabs_BPASS} quantify the actual radiation momentum rate absorbed by the medium by showing the cumulative fraction of this momentum that is absorbed relative to the incident radiation momentum. Since the outcome depends sensitively on metallicity, results are presented as a function of stellar age and gas density, for two characteristic temperature-density tracks: one following the equilibrium cooling curve at solar metallicity (middle panel), and the other corresponding to primordial composition gas (bottom panel). For reference, the two tracks are shown in the top panel, with coloured dots indicating the corresponding density displayed in the lower panels. 

Fig.~\ref{fig:fabs_BPASS} illustrates that the efficient transfer of radiation momentum from an SSP to the gas occurs primarily in dense (cold) gas, rich in neutral and molecular hydrogen and dust. At lower densities, and consequently higher temperatures ($T\gtrsim 10^4$ K), dust is destroyed and the gas becomes highly ionized, reducing its opacity and weakening the coupling between radiation and matter. This coupling is further complicated by a time-dependent interplay between the evolving stellar energy distribution (SED) and the metallicity-dependent opacity of the gas. At solar metallicity, the dense gas is already highly opaque, leading to a consistently high absorption efficiency that declines slightly over time as the stellar population's SED softens. In contrast, at primordial metallicity, the initially hard SED from hot, metal-free stars is effectively absorbed, but the aging population's SED cannot provide enough photons that couple to the highly ionized transparent metal-poor gas, causing the momentum coupling efficiency to drop over time.

\begin{figure}
\includegraphics[width=\columnwidth]{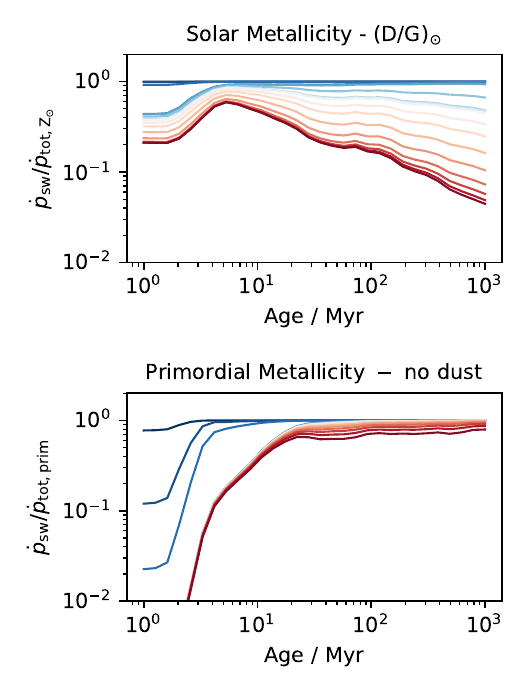}
    \caption{\textit{Top:} Fraction of specific stellar winds-driven momentum rate relative to the total effective momentum rate absorbed, as a function of stellar age (x-axis) for solar metallicity gas and a solar metallicity SSP. Different colours correspond to the densities and temperatures marked in the top panel of Fig.~\ref{fig:fabs_BPASS}. \textit{Bottom:} Same as the middle panel, but for primordial metallicity gas.}
\label{Fig:fraction_of_winds}
\end{figure}

Finally, Fig.~\ref{Fig:fraction_of_winds} shows the fraction of stellar wind–driven momentum rate relative to the total effective absorbed momentum rate, for both solar (top) and primordial (bottom) metallicity gas. The total effective momentum rate here refers not to the full momentum budget predicted by {\tt BPASS} (as shown in Fig.~\ref{fig:momentum_stellar_winds}), but rather to the combined contribution of stellar winds and the fraction of radiation momentum that is actually absorbed by the gas in our model. Results are presented for a range of densities and temperatures along the equilibrium cooling curve tracks displayed in the top panel of Fig.~\ref{fig:fabs_BPASS}, with line colours corresponding to the gas density of the dots in that panel. Fig.~\ref{Fig:fraction_of_winds} highlights that although radiation pressure dominates the intrinsic momentum budget, its inefficient coupling to the gas---especially at low densities---makes stellar winds the dominant source of absorbed momentum in diffuse ($n_{\rm H} \lesssim 0.1 \rm \ cm^{-3}$) environments. At typical ISM densities ($n_{\rm H} \gtrsim 1 \rm \ cm^{-3}$) and for solar metallicity, winds contribute roughly half of the absorbed momentum at early times, with radiation pressure accounting for the remainder. For primordial composition gas, stellar winds contribute very little during the early stages, making radiation pressure the dominant source of early feedback.

Having established the relative contributions of stellar winds and radiation pressure, we now turn to the physical rationale underlying the formation of \ion{H}{II} regions in our NEPS model.

\subsection{\ion{H}{II} regions around young stellar populations}

Radiation from young massive stars produces ionized regions (\ion{H}{II} regions), with characteristic temperatures of $T \sim 10^4$ K, resulting from the balance between photoheating and radiative cooling. In a homogeneous medium illuminated by a single ionizing source and assuming photoionization equilibrium, the size of an \ion{H}{II} region is given by the Str\"omgren radius:
\begin{align}\label{eq:stromgrenR}
        r_{\ion{H}{II}} &=\left ( \frac{3\dot Q(t) m_{*} }{4\pi\alpha_{\mathrm{B}}n_{\mathrm{H}}^2} \right )^{1/3},
\end{align}
where $\dot Q (t)$ is the ionizing photon rate per unit stellar mass, $m_{*}$ is the stellar mass of the source, $n_{\rm H}$ is the hydrogen number density of the medium, and $\alpha_{\rm B}$ is the case-B recombination coefficient. For a temperature of $T=10^4$ K, $\alpha_{\mathrm{B}} \approx 2.6\times10^{-13}\,\mathrm{cm}^{3}\,\mathrm{s}^{-1}$, yielding a Str\"omgren radius of:
\begin{equation}
    \label{eq:stromgrenR_physical}
    r_{\ion{H}{II}} \approx 0.46 \,\mathrm{kpc} \left ( \frac{\dot{Q}(t)}{10^{13}\,\mathrm{g}^{-1}\mathrm{s}^{-1}} \right )^{1/3} \left ( \frac{m_{*}}{10^{5}\,\mathrm{M}_{\odot}}\right )^{1/3} \left ( \frac{n_{\mathrm{H}}}{1\,\mathrm{cm}^{-3}} \right)^{-2/3}.
\end{equation}
Note that the development of the ionized cavity is not instantaneous; its size depends on time. The time-dependent size of the~\ion{H}{II} region can be written as:
\begin{equation}\label{eq:stromgrenRtdep}
    r_{\ion{H}{II}} (t) = r_{\ion{H}{II}} \left (1 - e^{-t/t_{\mathrm{rec}}} \right )^{1/3},
\end{equation}
where the recombination timescale, $t_{\rm rec}=(\nH \alpha_{\rm B})^{-1}$, for gas at $T=10^4$ K, becomes:
\begin{equation}
    t_{\rm rec} \approx 0.1 \ {\rm Myr} \left ( \displaystyle\frac{\nH }{1 \rm \ cm^{-3}} \right )^{-1}.
    \label{Eq:recombination}
\end{equation}

The previous equation shows that, at typical ISM densities, the formation timescale of ionized \ion{H}{II} cavities of size $r_{\ion{H}{II}}$ is much shorter than the lifetimes of young massive stars. In our model we compute the time-dependent evolution of $r_{\ion{H}{II}}$ self-consistently using Equation~\ref{eq:stromgrenRtdep}.

\begin{figure}
\includegraphics[width=\columnwidth]{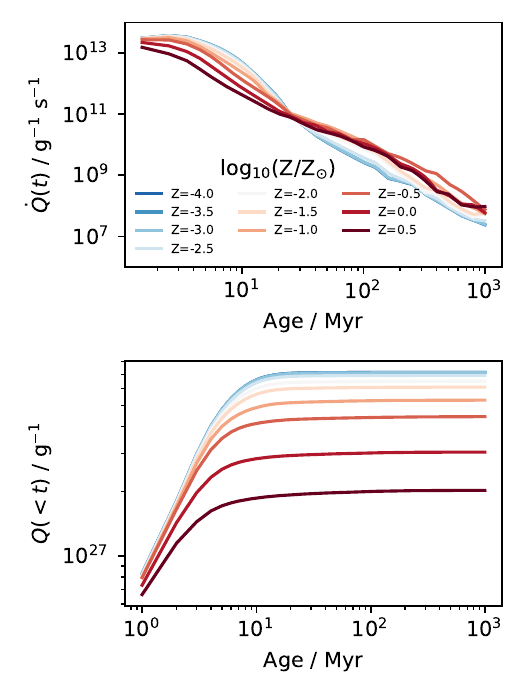}
    \caption{Ionizing photon rate (top) and cumulative number of ionizing photons (bottom), emitted per unit mass, as a function of stellar age, for SSPs of different metallicities (coloured lines), as predicted by {\tt BPASS}.}
\label{Fig:Qion_BPASS}
\end{figure}

Fig.~\ref{Fig:Qion_BPASS} presents the ionizing photon emission rate (top) and the cumulative number of ionizing photons (bottom) per unit stellar mass, as predicted by {\tt BPASS}. The rapidly declining emission rate implies that the majority of photons are released within the first $10 \ \mathrm{Myr}$ after star formation. Both the instantaneous rate and the integrated photon budget depend strongly on metallicity, with low-metallicity SSPs producing substantially more ionizing photons, and at higher rates, than their metal-rich counterparts.

\subsection{Numerical implementation}

Having outlined the physical processes and approximations underlying our simple model for stellar winds, radiation pressure, photoionization, and photoheating, we now describe how these early feedback channels are implemented in the {\tt COLIBRE} model of galaxy formation~\citep{Schaye2025}, which uses the {\tt SWIFT} Smoothed Particle Hydrodynamics (SPH) numerical code~\citep{Schaller2024}.

\subsubsection{Momentum injection}

In an SPH simulation, directly injecting momentum into dense gas particles is prone to rapid dissipation: because these particles are tightly packed and have short hydrodynamical timescales, the injected kinetic energy quickly thermalizes and is redistributed to neighboring particles through pressure forces and shocks, preventing the feedback from driving significant gas motions. Moreover, radiation pressure couples efficiently only to the densest gas, further limiting the impact of kinetic momentum injection. To circumvent these potential issues, we adopt a stochastic momentum injection scheme similar to that of~\cite{DallaVecchia2008}, in which the available momentum is delivered in discrete kicks with controlled velocities, ensuring that the total momentum budget is conserved while allowing individual particles to respond dynamically before dissipation dominates. These kicks are set to have a large velocity relative to the medium sound speed, which effectively causes the total available momentum to be distributed only to a subset of particles of the star SPH kernel. Note that here we refer to the SPH kernel of a stellar particle, rather than that of a gas particle. Although stellar particles are collisionless in {\tt COLIBRE}, the code assigns them an effective SPH kernel that defines their domain of influence based on the surrounding gas particles. This kernel is used both for injecting stellar feedback and for continuously distributing metals into the surrounding gas. We leverage this infrastructure for our stochastic NEPS momentum injection implementation.

Within the stochastic implementation approach, if we aim for a velocity kick, $\Delta v_{0}$, then the probability of giving an SPH gas particle this velocity kick in a time step, $\Delta t$, is:
\begin{equation}
    \label{Eq:current_implementation}
    {\mathcal P} \left (\Delta v_{0} \right ) =  \dot p_{*}(t_{*}, Z_{*}) \left ( \displaystyle\frac{\Delta t}{\Delta v_{0}} \right ) \left ( \displaystyle\frac{m_{*}}{M_{{\rm ngb}}} \right ),
\end{equation}
where $m_{*}$ is the mass of the newly formed stellar particle (representing an entire SSP), and $M_{\rm ngb}$ is the mass contained within the stellar particle SPH kernel. We can approximate $M_{{\rm ngb}}$ by $M_{{\rm ngb}} \approx N_{\rm ngb} m_{{\rm g}}$, where $N_{\rm ngb}$ is the typical number of gas particles within the kernel, and $m_{{\rm g}}$ is the gas particle mass. Since the gas and star particle masses are nearly identical in {\tt COLIBRE}, the probabilities are largely insensitive to the numerical resolution. Equation~\ref{Eq:current_implementation} shows that the probability of kicking a gas particle decreases inversely with the chosen velocity kick, $\Delta v_0$: stronger kicks imply that fewer particles can be selected for momentum injection, while weaker kicks distribute the available momentum across more neighbors. 

According to {\tt BPASS}, the total specific momentum rate that a young SSP can impart to its surroundings typically does not exceed $\dot p \approx 3 \times 10^{6} \ \rm cm \ s^{-1} \  Myr^{-1}$. For a desired velocity kick of $\Delta v_0 = 50 \ \rm km \ s^{-1}$, the probability of a momentum injection event affecting a gas particle is therefore below 10\% over the $\sim 10 \ \rm Myr$ lifetime of a massive O star, assuming $N_{\rm ngb}=100$ neighbors for simplicity. Consequently, momentum injection at this kick velocity would, on average, directly affect less than 10\% of a collapsing self-gravitating SPH cloud within 10 Myr. In practice, the SPH kernel may contain fewer particles, which would double this efficiency; for instance, assuming 50 neighbors---an unrealistically low bound for the {\tt COLIBRE} SPH formulation---would raise the maximum effective momentum coupling to $\sim 20$\%. Note that despite this relatively weak coupling, particles receiving these kicks may, in turn, disturb their neighbors, potentially generating turbulence within the collapsing cloud. 

Alternatively, enforcing a probability of unity for all particles within the SPH kernel yields a velocity kick $\Delta v_0 \approx 3 \rm \ km \ s^{-1}$, which is so small that it is not expected to affect significantly the collapse of large gas clouds forming in simulations that cannot resolve the internal structure of cold molecular clouds. 

In {\tt COLIBRE}, we adopt $\Delta v_{0} = 50\ \rm km\ s^{-1}$, a compromise between a velocity large enough to disperse a small fraction of the gas particles contributing to the formation of massive gas clouds in the ISM, and a value that is physically reasonable, being highly supersonic relative to the cold and warm ISM. For reference, dispersing a minimal\footnote{A minimal SPH gas cloud is a collection of $N_{\rm ngb}$ particles contained within the SPH kernel's domain. It is the smallest gravitationally-bound gas structure than can be ``resolved'' in the simulation.} self-gravitating SPH cloud requires kicking its particles with velocities significantly exceeding the cloud’s sound speed. In {\tt COLIBRE}, star formation depends on the local gas density and temperature, but stars predominantly form in gas colder than $\sim 1000\ \rm K$, corresponding to a sound speed of $\approx 4\ \rm km\ s^{-1}$. A velocity kick of 50 km s$^{-1}$ is thus highly supersonic relative to the cold ISM in which stars form, and also supersonic relative to the warm ($T\sim 10^4$ K) ISM. For reference, the adopted velocity kick for supernova-driven turbulence in {\tt COLIBRE} matches our adopted value for $\Delta v_{0}$~\citep{Chaikin2023}.

We explore the impact of varying this parameter in Appendix~\ref{Ap:kick_velocity}, where we demonstrate that for momentum to be effective at suppressing runaway collapse and regulating the star formation history of an unstable disk, the specific value of the kick velocity needs to be highly supersonic, with velocities exceeding $\sim 20 \rm \ km \ s^{-1}$. For stable disks, the exact choice of the kick velocity is largely irrelevant. Furthermore, as we show in Section~\ref{Sec:early_feedback}, thermal feedback from~\ion{H}{II} regions is the dominant regulatory process within our NEPS feedback model. 

\subsubsection{\ion{H}{II} regions}
\label{Sec:HIIregions-implementation}

While precise treatment of \ion{H}{II} regions requires multi-frequency radiation hydrodynamics and very high numerical resolution, an approximate treatment can capture its main effects. Given the numerical complexity and computational cost of a full treatment and the need to define a physical search radius around young stars to deposit ionizing photons, a sensible compromise is to limit the growth of \ion{H}{II} regions to the minimum scale of self-gravitating SPH clumps where stars form. This approach allows the onset of ionized and photoheated regions to be modeled approximately, without the high computational cost of depositing photons beyond the SPH kernel of stars, while taking full advantage of the existing SPH framework already in place in {\tt SWIFT}. For sufficiently high densities, the Strömgren radius is entirely contained within the SPH kernel of the stars, making this approximation reasonable. 

We can determine the density above which this approximation is valid by requiring that the mass of an \ion{H}{II} region,
\begin{equation}
M_{\ion{H}{II}} = \rho_{\rm gas} \displaystyle\frac{4}{3}\pi r_{\ion{H}{II}}^3,
\end{equation}
does not exceed the minimum mass of an SPH self-bound clump, $M_{\rm ngb}$. Equating these two quantities defines a threshold gas density, $\rho_{\rm gas}$, above which the Strömgren radius remains fully contained within the SPH kernel, and below which the ionized region would extend beyond the kernel. Expressed in terms of the hydrogen number density,~$n_{\rm H}=\rho_{\rm gas} X_{\rm H}/m_{\rm p}$, this threshold is:
\begin{equation}
\nH \gtrsim n_{\rm H,min} = 1 \ {\rm cm^{-3}} \left ( \frac{\dot{Q}(t)}{10^{13}\,\mathrm{g}^{-1}\mathrm{s}^{-1}} \right ) \left ( \frac{m_{*}}{m_{\rm g}}\right ) \left ( \displaystyle\frac{X_{\rm H}}{0.75} \right )^{-1} \left ( \frac{N_{\rm ngb}}{100} \right )^{-1},
\label{Eq:NHmin}
\end{equation}
where $X_{\rm H}$ is the hydrogen mass fraction. This minimum density threshold, $n_{\rm H,min}$, is insensitive to the adopted numerical resolution if the gas particle mass is similar to that of the star particles. However, although for densities above $n_{\rm H,min}$ there is sufficient mass within the SPH kernel to absorb the ionizing photons emitted by a newly formed SSP, the methodology eventually breaks down at very high densities. As the gas density increases, the physical mass of an \ion{H}{II} region decreases, eventually becoming smaller than that of a single SPH gas particle. This limitation is evident from the \ion{H}{II} mass definition, which can also be expressed explicitly as a function of density:
\begin{equation}
    M_{\ion{H}{II}} \approx 10^{8} \ {\rm M_{\odot}} \left ( \frac{\dot{Q}(t)}{10^{13}\,\mathrm{g}^{-1}\mathrm{s}^{-1}} \right ) \left ( \frac{m_{*}}{10^{6}\,\mathrm{M}_{\odot}}\right ) \left ( \frac{n_{\mathrm{H}}}{1\,\mathrm{cm}^{-3}}\right)^{-1} \left ( \displaystyle\frac{X_{\rm H}}{0.75} \right )^{-1}.
    \label{Eq:HII-mass}
\end{equation}
Equating $M_{\ion{H}{II}} = m_{\rm g}$ defines the density threshold above which the mass of a single SPH particle exceeds that of an \ion{H}{II} region.
\begin{equation}
    \nH \gtrsim n_{\rm H,max} \approx 200 \ {\rm cm^{-3}} \left ( \frac{\dot{Q}(t)}{10^{13}\,\mathrm{g}^{-1}\mathrm{s}^{-1}} \right ) \left ( \frac{m_{*}}{m_{g}}\right ) \left ( \displaystyle\frac{X_{\rm H}}{0.75} \right )^{-1}.
\end{equation}
Similarly to $n_{\rm H, min}$, the threshold density $n_{\rm H, max}$ is independent of numerical resolution, since it depends on the ratio of stellar to gas particle masses, which are approximately equal in {\tt COLIBRE}. 

Thus, the development of \ion{H}{II} regions around young SSPs requires that the mass of the ionized cavity does not exceed that of the SPH kernel. Otherwise, an SSP forming inside the cloud would be unable to deposit its ionizing photons, leading to energy losses. Conversely, in an SPH simulation, the mass of the \ion{H}{II} region cannot fall below that of a single gas particle.

To mitigate these issues we implement the heating and ionization of individual gas particles in a stochastic manner, ensuring that the total mass of \ion{H}{II} regions is correct on average. This is accomplished by assigning each gas particle in the vicinity of a young SSP---i.e., within the SPH kernel of a newly formed star particle---the following probability of being tagged as part of an \ion{H}{II} region:
\begin{equation}
 {\mathcal P_{\ion{H}{II}}} = {\rm min} \left ( \displaystyle\frac{M_{\ion{H}{II}}}{M_{\rm ngb}}, 1\right ).
 \label{Eq:HII_region}
\end{equation}
This probability is insensitive to the simulation resolution, provided that the masses of gas and star particles are comparable, which is the case in {\tt COLIBRE}.

For gas particles labelled as \ion{H}{II} regions we perform the following operations:
\begin{enumerate}
    \item we set their temperature to $T=10^4$ K;
    \item we fully ionize their hydrogen and helium;
    \item we prevent them from cooling below $10^{4}$ K, recombining, and forming stars during a time interval $\Delta t_{\ion{H}{II}}$, after which their temperature and ion abundances can evolve freely.
\end{enumerate}

Choosing a small value for $\Delta t_{\ion{H}{II}}$ ensures good temporal sampling of the SSP's radiation field and prevents star particles from prematurely leaving their own~\ion{H}{II} regions. We therefore recalculate~\ion{H}{II} regions around young star particles at discrete time intervals, $\Delta t_{\ion{H}{II}} = 2$ Myr. At each step, the mass of the \ion{H}{II} region is calculated based on the average ionizing luminosity the SSP will emit. The gas particles within the star particle's SPH kernel are then stochastically selected to be part of the~\ion{H}{II} region based on the probability given by Eq.~\ref{Eq:HII_region}, and prevented from recombining and cooling during $\Delta t_{\ion{H}{II}}$. This process continues until the star particle is older than 50 Myr, by which time the vast majority of ionizing photons have been emitted. After a time $\Delta t_{\ion{H}{II}}$, the~\ion{H}{II} region tag is removed, and the temperature and ion abundances of the gas particles can evolve again freely according to the cooling and chemistry solver. 

We stress that the choice of $\Delta t_{\ion{H}{II}}=2$ Myr ensures that the thermal energy has a lasting dynamical impact. At high densities ($n_{\rm H} \gg 1$ cm$^{-3}$), where star formation occurs, the physical recombination and cooling times are typically shorter than 2 Myr. If we were to allow the gas to cool and recombine according to the physical rates, the injected thermal energy would be radiated away almost instantaneously, before it could exert significant pressure and have a dynamical impact. This issue is a common challenge for feedback implementations at finite resolution. We explicitly avoid scaling $\Delta t_{\ion{H}{II}}$ with the local density, as this would dramatically reduce the feedback duration in the densest gas, introducing numerical overcooling. Instead, the choice of $\Delta t_{\ion{H}{II}} = 2$ Myr ensures the heating duration is comparable to the sound-crossing time of the star-forming gas in our high-resolution simulations, allowing pressure to effectively drive expansion. We explore the sensitivity of our model to the value of $\Delta t_{\ion{H}{II}}$ in Appendix~\ref{Ap:HII_time}. As shown in Fig.~\ref{fig:HII_time}, choosing values smaller than $1$ Myr leads to overcooling, rendering the feedback ineffective. A value of $\Delta t_{\ion{H}{II}} \gtrsim 1$ Myr is required to ensure the thermal energy couples dynamically to the gas.

Our implementation is therefore functionally a ``delayed cooling'' scheme, which artificially sustains a high-pressure, photoionized phase around young stars for a duration comparable to the early lifetime of massive stars, until the onset of core-collapse SNe. This approach ensures that the thermal energy budget from photoionization is effectively coupled to the gas as mechanical work, providing the pressure support needed to regulate local star formation prior to the first core-collapse SN events begin at roughly 3 Myr (see the dashed line in the bottom panel of Fig.~\ref{fig:momentum_stellar_winds}). This method should be regarded as a subgrid model for the integrated pressure exerted by an \ion{H}{II} region over its early lifetime, rather than as a direct simulation of the ionization–recombination cycle around star-forming regions.

Although our model captures the local pressurization from ~\ion{H}{II} regions, a key limitation is that feedback is confined to the SPH kernel of individual star particles. As a result, the total mass of overlapping ~\ion{H}{II} regions may be underestimated when multiple star particles influence the same gas.

Finally, we note that for gas densities significantly lower than those of the ISM ($n_{\rm H} \ll 1 \ \mathrm{cm^{-3}}$; Equation~\ref{Eq:NHmin}), our model underestimates the amount of material photoionized by newly formed stars. We acknowledge this inconsistency on these scales, but we note that this is unlikely to impact the properties of simulated galaxies since (i) star formation typically occurs at higher densities unless the metallicity is very high or the resolution too poor; and (ii) at these low densities, competing effects---such as photoionization and photoheating from the collective interstellar radiation field and the external UVB--dominate the photon budget, regardless of the star formation prescription. Both of these processes are modeled in {\tt COLIBRE}---so the additional heating from our NEPS module would be hydrodynamically unimportant.

\section{Numerical code and galaxy formation model}
\label{Sec:Code}
As we will explain later on, to test the implementation of our {\tt COLIBRE} NEPS feedback module, we carry out a suite of idealized disk galaxy simulations, described in detail in Section~\ref{Sec:Instabilities}. Compared to large-volume simulations starting from cosmological initial conditions, these controlled experiments offer two key advantages: they are computationally less expensive, and they allow us to isolate the impact of early feedback in a systematic manner. The application of the {\tt COLIBRE} galaxy formation model---including the NEPS feedback alongside other baryonic physics relevant to galaxy formation---is presented in detail by~\cite{Schaye2025}.

\subsection{COLIBRE galaxy formation model}

The {\tt COLIBRE} galaxy formation model is implemented within the open-source N-body+SPH code {\tt SWIFT}~\citep{Schaller2024}. Here we summarize the {\tt COLIBRE} subgrid prescriptions that operate on top of {\tt SWIFT}’s gravity+hydrodynamics solver---specifically those governing gas cooling, star formation, and SN feedback energy injection---which are most relevant for the simulations presented in this work.

On top of the modules described below, {\tt COLIBRE} also includes the following processes: stellar mass loss and chemical enrichment, turbulent diffusion, dust grains, supermassive black holes, and AGN feedback. These processes, although crucial for galaxy formation in large-volume simulations, are less relevant for our purposes, so we refer the reader to~\cite{Schaye2025} for a discussion of the full model.

\subsubsection{Radiative cooling and heating}

Radiative cooling and heating are computed as described in~\cite{Ploeckinger2025}. In short, these rates are calculated using the {\tt HYBRID-CHIMES} module, which tracks cooling down to temperatures of approximately $10$ K, a process made possible by the explicit tracking of molecular hydrogen, dust cooling and heating, metal-line cooling, and shielding by dust and gas. A key feature of the cooling model is its hybrid approach to chemical evolution. The abundances of nine hydrogen and helium species, including molecular hydrogen, are tracked in non-equilibrium using the {\tt CHIMES} chemical network~\citep{Richings2014a, Richings2014b}. In contrast, for computational efficiency, the cooling rates for 9 metal coolants (C, N, O, Ne, Mg, Si, S, Ca, and Fe) are precomputed using {\tt CHIMES} assuming chemical equilibrium. Crucially, however, these metal line cooling rates are corrected on-the-fly using the nonequilibrium free electron density provided by the hydrogen and helium atoms, ensuring a more self-consistent treatment of cooling and heating than simulations assuming ionization equilibrium.

The cooling and chemistry are coupled to the on-the-fly model for dust grain evolution, presented in~\cite{Trayford2025}. This coupling allows the simulation to self-consistently account for several critical processes, including the depletion of gas-phase metals onto dust grains, which direcly affects metal-line cooling; the formation of molecular hydrogen on the surface of dust grains; and photoelectric heating by dust grains and direct dust cooling from grain-to-gas collisions, with rates that depend on the grain size distribution.

While the simulations do not perform full radiative transfer, they incorporate its main effects through an approximate treatment of shielding. The attenuation of the background radiation from the UVB and from local stellar sources by gas and dust is estimated using the local Jeans column density. This approximation, strictly valid only for self-gravitating gas, accounts for shielding by dust, \ion{H}{I}, \ion{H$_2$}, \ion{He}{I}, \ion{He}{II}, and CO, and uses the maximum of the thermal and turbulent Jeans lengths to account for possible effects of turbulence in the simulation. 

\begin{figure}
    \includegraphics[width=\columnwidth]{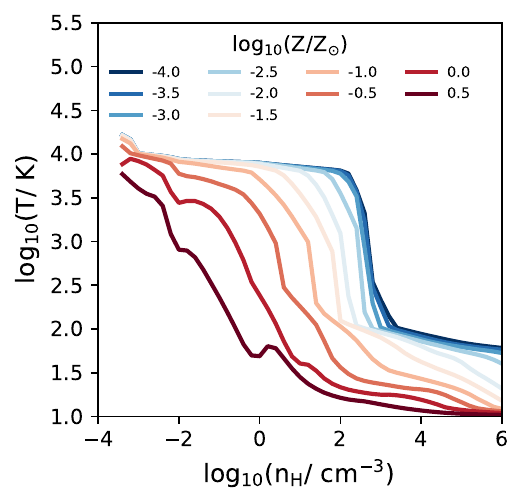}
    \caption{Redshift $z=0$ {\tt COLIBRE} equilibrium temperature as a function of density in ionization equilibrium, for various metallicities.}
    \label{Fig:equilibrium-curves}
\end{figure}

We note that for the numerical experiments presented in this work, we use precomputed cooling tables that assume ionization equilibrium for numerical efficiency. The full non-equilibrium chemistry and on-the-fly dust models are therefore not activated. For reference, Fig.~\ref{Fig:equilibrium-curves} shows typical density-temperature tracks for gas in ionization equilibrium resulting from these tables, coloured by metallicity. These tracks correspond to the equilibrium temperatures at which photoheating---due to the external UVB and the adopted interstellar radiation field---balances radiative cooling. 

\subsubsection{Star formation}

In {\tt COLIBRE}, stars form in gas that is locally self-gravitating, following the prescription presented by~\cite{Nobels2024}. In short, star formation is implemented stochastically, following the Schmidt law~\citep{Schmidt1959} with an efficiency per free fall time of $1\%$, and is computed on gas particles that are locally gravitationally unstable. The gravitational instability criterion is assessed by requiring that the local free-fall time of a minimal SPH gas cloud is shorter than the sound crossing time over the SPH kernel, for which the ``effective'' sound speed is calculated by adding in quadrature the thermal and tubulent velocity dispersions. This requirement makes gas particles eligible to form stars provided:
\begin{equation}
\displaystyle\frac{\sigma_{\rm therm}^2 + \sigma_{\rm turb,3D}^2}{G N_{\rm ngb}^{2/3} m_{\rm g,i}^{2/3} \rho_{\rm gas}^{1/3}} < 1
\end{equation}
In the previous equation, $\sigma_{\rm therm}$ and $\sigma_{\rm turb,3D}$ are the thermal and 3D turbulent velocity dispersion, respectively; $m_{\rm g}$ and $\rho_{\rm gas}$ are, respectively, the mass of the $i$-th gas particle and its density, and $G$ is the gravitational constant.

\begin{figure*}
    \includegraphics[]{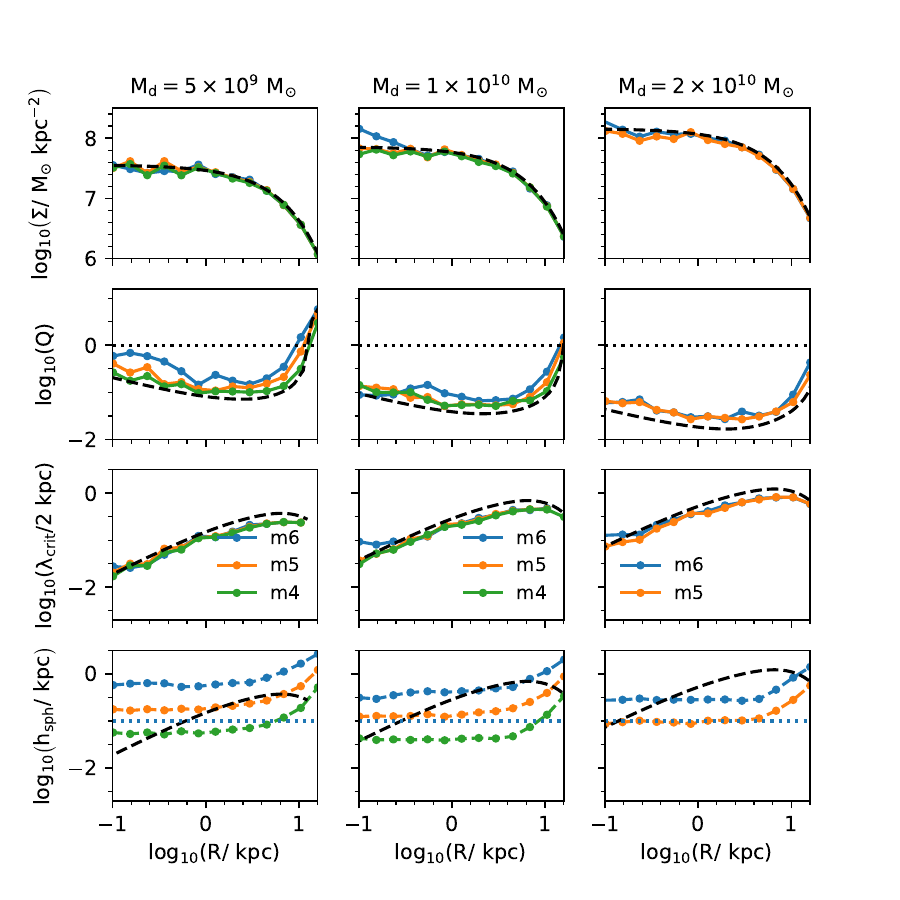}
    \caption{Stability analysis of the gaseous disks considered for our numerical experiments. From left to right, each column varies the initial mass of the exponential gas disk by factors of 2, from $5 \times 10^9 \ \rm M_{\odot}$ to $2 \times 10^{10} \rm \ M_{\odot}$. The rows show, from top to bottom respectively, and as a function of radius: the gaseous disk surface density, its Toomre parameter $Q$, the most unstable wavelength, and the disk median SPH smoothing length. The dashed lines indicate the values used to initialize the disk numerically, whereas the colored lines (dots) indicate the values measured in the simulation after evolving the disk for $100 \ \rm Myr$, including star formation but without feedback processes, varying the numerical resolution, with the blue line indicating a gas particle mass of $10^6 \rm \ M_{\odot}$, the orange line corresponding to $10^5 \rm \ M_{\odot}$, and the green line to $10^4 \rm \ M_{\odot}$. The horizontal dotted line in the bottom panels indicates the gravitational softening length. For comparison, the dashed black lines in the bottom row indicate $\lambda_{\rm crit}/2$.}
    \label{fig:Radial_instabilities_ICs}
\end{figure*}

\subsubsection{Core-collapse supernova feedback}

{\tt COLIBRE} includes a SN feedback model designed to drive large-scale galactic winds. The scheme builds on the stochastic thermal feedback method of \citet{DallaVecchia2012}, in which the energy from core-collapse SNe is accumulated and released in discrete heating events. In each event, neighbouring gas particles are heated by a large temperature increment, $\Delta T_{\rm SN}$, ensuring that the injected energy is not immediately radiated away.

Relative to previous implementations, {\tt COLIBRE} introduces several key innovations. First, energy by core-collapse SNe is injected in a dual mode: the bulk ($90$\%) is delivered thermally, while a smaller fraction is imparted kinetically through low-velocity ($50$ km s$^{-1}$) kicks to neighbouring gas particles using the implementation of~\cite{Chaikin2023}. Second, the temperature increment, $\Delta T_{\rm SN}$, is not fixed but increases with ambient gas density, allowing SN events to more effectively target lower density material while mitigating radiative losses at higher densities. Third, the energy budget available for feedback is not constant, but increases with the thermal pressure of the cloud from which stars are born, making SN feedback naturally more effective in high-pressure environments. Fourth, the model also accounts for Type Ia SNe, which contribute energy through the same stochastic thermal scheme but without the pressure dependence. Finally, the {\tt COLIBRE} feedback scheme distributes the energy to the gas particles in a statistically isotropic manner, following the algorithm of~\cite{Chaikin2022}.

For a complete description of the {\tt COLIBRE} SN feedback model, we refer the reader to~\cite{Schaye2025}. In this work, we include SN feedback from core-collapse and type-Ia SNe, setting the main parameters to the following values: $\Delta T_{\rm min} = 10^{6.75}$ K, $\Delta T_{\rm max} = 10^{8}$ K, $n_{\rm H, pivot} = 0.6 \rm \ cm^{-3}$, $f_{\rm E, min} = 0.3$, $f_{\rm E, max} = 4.0$, and $P_{\rm E, pivot}/k_{\rm B}=1.5\times 10^4 \rm \ K \ cm^{-3}$, independently of resolution~\citep[see][for a description of these parameters]{Schaye2025,Chaikin2025}. 

\section{The simulations}
\label{Sec:Simulations}

Having laid out the physical rationale and numerical implementation of our NEPS {\tt COLIBRE} module, we now proceed to test its impact in numerical simulations. We begin by describing the numerical setup. We note that while {\tt COLIBRE} explicitly models chemical enrichment and dust grain evolution, these subgrid modules have been deactivated for the numerical experiments discussed here.

\subsection{Initial conditions}
\label{Sec:Initial-conditions}
To test the implementation of our NEPS feedback module in {\tt COLIBRE}, we consider redshift $z=0$ simulations of isolated gaseous disks embedded in an external Navarro-Frenk-White dark matter halo~\citep{Navarro1996} of virial\footnote{We define virial quantities as those measured within a sphere where the mean density is 200 times the critical density of the Universe.} mass, $M_{200}=1.5 \times 10^{12} \ \rm M_{\odot}$, and concentration, $c=10$. To mimic the presence of an already formed thick stellar disk, we also include an external axisymmetric Miyamoto-Nagai potential~\citep[][MN hereafter]{Miyamoto1975} with a mass, $M_{\rm MN}=3 \times 10^{10} \rm \  M_{\odot}$, a scale radius, $R_{\rm MN}=4 \rm \ kpc$, and a scale height, $Z_{\rm MN}= 0.4 \rm \ kpc$.

The live gaseous component is initialized as an exponential disk with an initial mass $M_{\rm d}$, which we vary by factors of 2 in the range $M_{\rm d} = 5 \times 10^{9} \rm \  M_{\odot}$ to $M_{\rm d} = 2 \times 10^{10} \rm \ M_{\odot}$. The disk has a scale radius $R_{\rm d}=0.02\,r_{200}$, where $r_{200}$ is the virial radius of the dark matter halo, and a vertical scale height $z_{\rm d}=0.1 R_{\rm d}$. Note that the gaseous disk scale height quickly relaxes into vertical equilibrium so our results are not sensitive to the initial choice of $z_{\rm d}$. 

The gas particles sampling the gaseous disk are initially placed in circular orbits, with a circular velocity matching the sum in quadrature of the circular velocities of the halo, the stellar disk, and the live gaseous disk at the particles' location. In all cases, the gaseous component contributes negligibly to the overall circular velocity, so the rotation curve is dominated by the external halo and the stellar disk potentials. The initial temperature of the gas particles is set to $T=10^{5}$ K, and they quickly cool down to the equilibrium temperature of solar metallicty depicted in the top panel of Fig.~\ref{fig:fabs_BPASS}.

For our numerical experiments, we consider three resolution levels, sampling the disk with gas particle masses of $10^6$, $10^5$, and $10^4 \rm \ M_{\odot}$. We refer to these resolutions as {\tt m6}, {\tt m5}, and {\tt m4}, respectively, following the {\tt COLIBRE} nomenclature.\footnote{Note that for the large volume {\tt COLIBRE} simulations suite, {\tt m5}, and {\tt m6} correspond to gas particle masses of $\approx 10^{5.36}$ and $10^{6.26} \rm \ M_{\odot}$, respectively.} This setup results in simulations that sample the disk with as few as 5,000 particles for the least massive disk at the lowest resolution and up to 2,000,000 particles for the most massive disk at the highest resolution. However, due to the high computational cost of running the most massive disks at the highest resolution, we restrict simulations of the most massive disk to the {\tt m6} and {\tt m5} resolutions. Consequently, the most massive disk is initially realized with at most 200,000 gas particles. 

Unlike the usual approach of scaling the gravitational softening with resolution, we choose to keep it constant, approximately matching the value that a high-resolution {\tt COLIBRE} simulation would adopt at {\tt m4} resolution. In {\tt COLIBRE}, the adopted gravitational softenings are $\epsilon_{\rm m6} = 700\ \rm pc$ and $\epsilon_{\rm m5} = 350 \rm pc$~\citep[see Table 2 of][]{Schaye2025}, which would imply $\epsilon_{\rm m4} \approx 175\ \rm pc$. We adopt a slightly smaller value, $\epsilon_{\rm m4} = 100\ \rm pc$, in all our runs. This choice ensures that any lack of convergence in our experiments reflects the available spatial dynamic range (i.e., the number of particles) rather than variations in the gravitational softening. We note that, at the time of writing, {\tt COLIBRE} has not yet been run at {\tt m4} resolution within a cosmological large volume.

We aim to characterize our NEPS feedback module in the context of an already-formed galaxy. To this end, we initialize the gaseous disk with solar metallicity ($Z_{\odot} = 0.013$), and the solar relative abundance pattern. The choice of metallicity at fixed disk mass influences the stability of the disk, since the cooling efficiency---and thus the thermal pressure---of the gas at a given density depends on its metallicity (see Fig.~\ref{Fig:equilibrium-curves}).
Conversely, at fixed metallicity, the mass of the gaseous disk regulates both the overall stability of the system and the characteristic wavelength of the instabilities that develop. 

Finally, note that because we do not use the {\tt COLIBRE} live dust model, the dust-to-gas ratio in our simulations is solar for gas of solar metallicity or zero for primordial gas.

\begin{figure*}
    \centering
    \subfloat[Face-on gas distribution]{
        \includegraphics[width=0.48\textwidth]{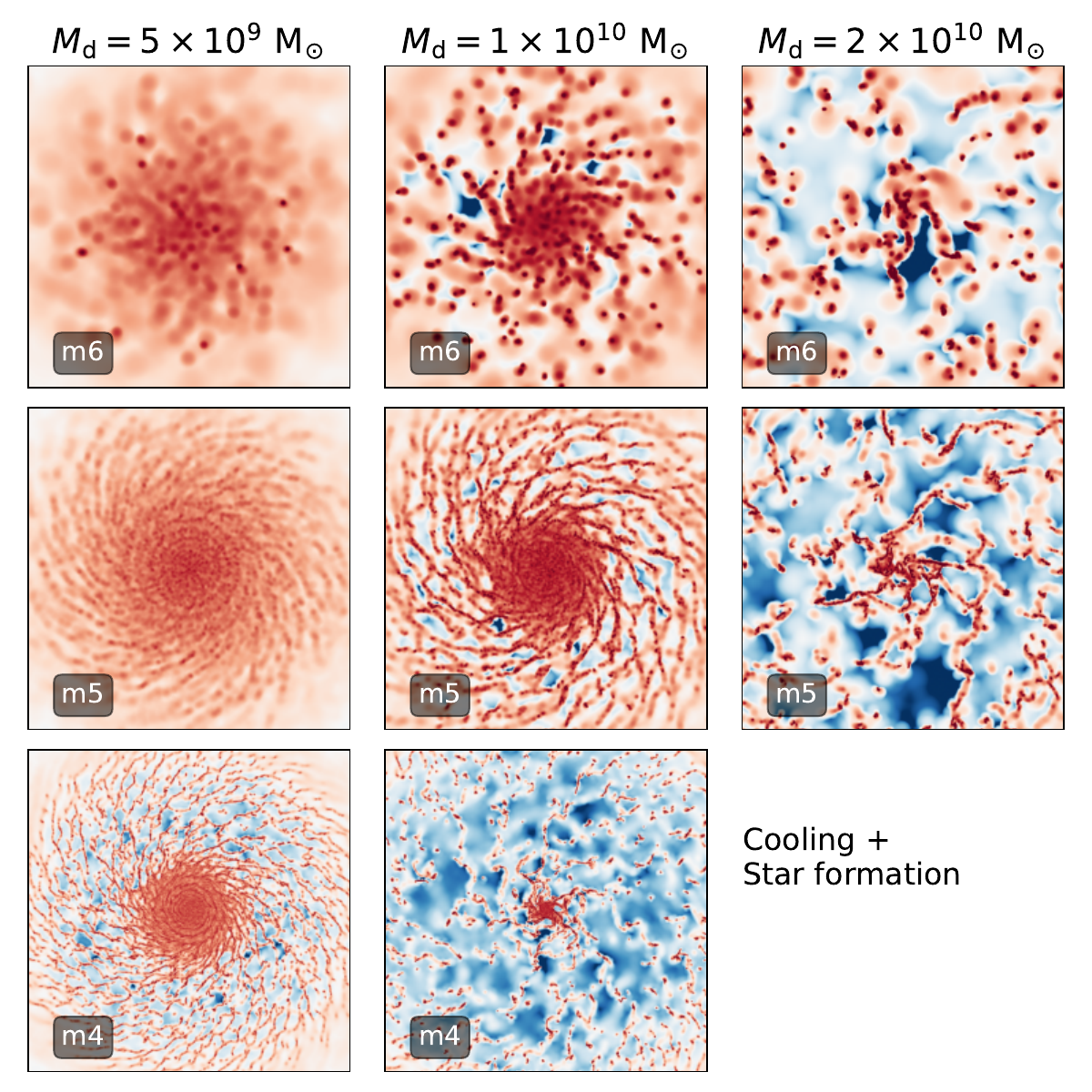}
        \label{fig:Gas-CoolingSFOnlye}
    }
    \hfill
    \subfloat[Face-on stellar distribution]{
        \includegraphics[width=0.48\textwidth]{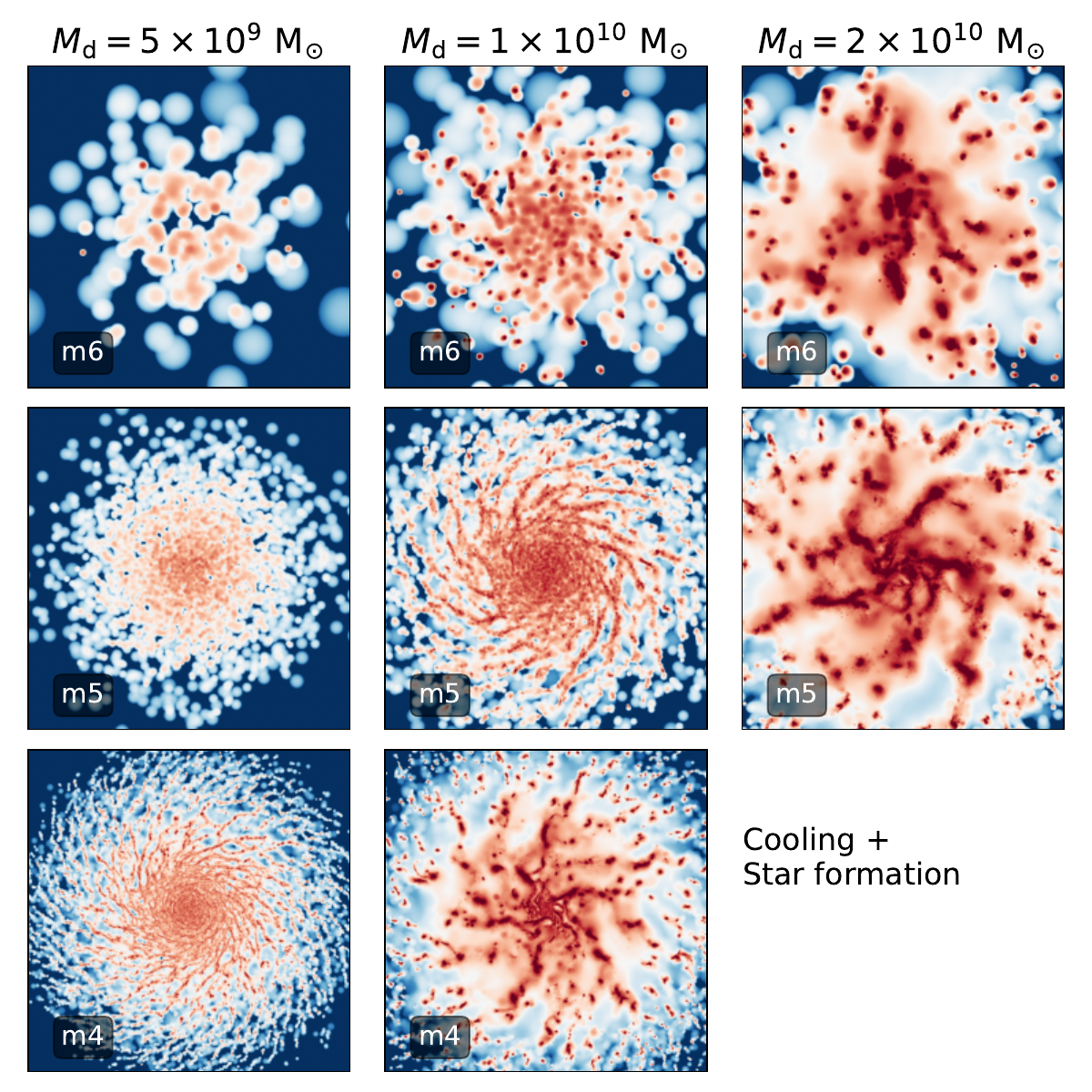}
        \label{fig:Stars-CoolingSFOnlye}
    }
    \caption{Face-on surface density maps of the gas (left) and stars (right) at $t=500 \rm \ Myr$. The simulations vary the initial gaseous disk mass, $M_{\rm d}$ (columns), and the gas particle mass resolution (rows). These runs include only gas cooling and star formation, while stellar feedback is neglected. The color map assigns bluer and redder colours to low- and high-density pixels respectively, and the minimum and maximum surface density values ($10^{4}$ and $10^8$ M$_{\odot}$ kpc$^{-2}$, respectively) are identical across the panels. The extent of the square boxes is $20 \rm \ kpc$. }
    \label{fig:SFOnly-combined}
\end{figure*}

\subsection{Intrinsic disk instabilities}
\label{Sec:Instabilities}

For the adopted metallicities and masses, all our disks are initially highly unstable according to the Toomre criterion~\citep[$Q < 1$;][]{Toomre1964}. The characteristic scale of the resulting instabilities is set primarily by the gas disk surface density. This follows from the fact that the gaseous component contributes negligibly to the rotation curve, so that the critical unstable wavelength, $\lambda_{\rm crit}$, depends only on the gas surface density, $\Sigma$:
\begin{equation}
    \lambda_{\rm crit} = \displaystyle\frac{4 \pi^2 G \Sigma}{\kappa^2},
    \label{Eq:lambda_crit}
\end{equation}
where the epicyclic frequency is given by $\kappa \approx 2 V_{\rm c}(R) / R$, which remains approximately constant across our disks. Here, $V_{\rm c}(R)$ denotes the total circular velocity at radius $R$.

Fig.~\ref{fig:Radial_instabilities_ICs} presents the stability analysis of the live gaseous disks. Each column corresponds to a different disk gas mass (indicated at the top), while the four rows display diagnostic quantities relevant for assessing disk stability. Coloured lines show direct measurements from simulations that include star formation but exclude feedback, with small dots marking the specific radii where measurements are performed. All quantities are evaluated after evolving the disks for $\approx 100 \rm \ Myr$, roughly one orbital time at $R_{\rm d}$. Results are shown for three numerical resolutions: $m_{\text{gas}} = 10^6 \, \text{M}_{\odot}$ ({\tt m6}, blue), $m_{\text{gas}} = 10^5 \, \text{M}_{\odot}$ ({\tt m5}, orange), and $m_{\text{gas}} = 10^4 \, \text{M}_{\odot}$ ({\tt m4}, green). For comparison, the black dashed lines indicate the values expected from the initial surface density profiles.

The top panels demonstrate that all disks remain in radial equilibrium, independent of resolution or mass : their profiles are unchanged relative to the initial conditions (dashed lines). 

The second row shows the Toomre parameter, $Q$, defined as:
\begin{equation}
    Q = \displaystyle\frac{c_{\rm s} \kappa}{\pi G \Sigma},
\end{equation}
where $c_{\rm s}$ is the mid-plane sound speed and $\kappa$ the epicyclic frequency. These panels confirm that, across all disk masses and resolutions, the systems are formally unstable ($Q<1$; horizontal dotted line), with the instability strengthening (i.e., $Q$ decreasing) as the disk gas mass increases. As we discuss below, however, the ability of a given simulation to actually capture these instabilities depends on resolution, and not all setups are able to do so.

The third row of Fig.~\ref{fig:Radial_instabilities_ICs} shows the most unstable wavelength, $\lambda_{\rm crit}/2$ (Equation~\ref{Eq:lambda_crit}), as a function of radius. The characteristic unstable scale increases with disk gas mass, ranging at $R=1 \rm \ kpc$ from a few hundred parsecs in the least massive disks to $\sim 1 \rm \ kpc$ in the most massive, with the largest wavelengths occurring at large radii.

The bottom row demonstrates that not all simulations can resolve these instabilities. Properly resolving an instability of wavelength $\lambda$ requires that both the SPH smoothing length\footnote{The hydrodynamical spatial resolution of an SPH simulation is largely controlled by particle smoothing lengths.} and the gravitational softening length be smaller than $\lambda$. The median smoothing lengths shown here exceed $\lambda_{\rm crit}/2$ unless either the resolution, the disk mass, or both are sufficiently high. For example, in the least massive disk (left column), only runs with $m_{\rm gas} < 10^5 \ M_{\odot}$ begin to marginally resolve unstable regions. By contrast, the {\tt m6} simulations cannot resolve $\lambda_{\rm crit}$ even with a few smoothing lengths and are therefore expected to be artificially stabilized everywhere unless the gas cools to lower temperatures and is therefore pushed to higher densities---an expectation we confirm later. In higher-mass disks, resolution determines which regions are stabilized, but in all cases some unstable regions are resolved.

Finally, we note that increasing the resolution by factors of ten reduces the median SPH smoothing length by a factor of $\approx 3$. This indicates that the disk’s volume density structure does not converge with resolution. If density were converged, increasing the mass resolution by a factor of ten would reduce the smoothing length by $10^{1/3} \approx 2.15$, smaller than what we observe. We trace this behavior to the extremely thin vertical structure of these cold disks, which remains unresolved in our simulations. Following~\cite{Benitez-Llambay2018}, we estimate the physical scale height of our disks to be well below a parsec---far smaller than the adopted gravitational softening---so the mid-plane density, $\rho \sim \Sigma / H$, cannot be expected to converge.

This underscores that resolving the vertical structure of cold disks with {\tt COLIBRE} requires extremely high resolution, beyond the scope of the present simulations. In simulations including feedback processes, the increased thermal and turbulent pressure is expected to make the convergence of the vertical scale-height less stringent with numerical resolution. Despite these limitations, we consider this suite suitable for testing the {\tt COLIBRE} NEPS module.

\section{Results}
\label{Sec:Results}

Starting from the initial conditions described in Section~\ref{Sec:Initial-conditions}, we evolve a simplified version of the {\tt COLIBRE} model to explore the impact of the NEPS feedback module. We begin by presenting a reference simulation that includes only gas cooling, heating from the UVB and the diffuse interstellar radiation field, and star formation, without any form of feedback. This provides a baseline against which we can assess the effects of early feedback on the evolution  and properties of the idealized disks. Later, and to isolate the impact of the NEPS feedback channel, we will consider simulations with NEPS feedback active while neglecting SN feedback altogether. This helps prevent strong outflows that could otherwise mask the influence of the less energetic early feedback processes. We study the interplay between the {\tt COLIBRE} SN feedback module and our NEPS channel in Section~\ref{Sec:pre-SN+Feedback}. All simulations share the same initial conditions.

\subsection{Simulations with gas cooling, heating, and star formation}

We begin by examining simulations that include only gas cooling, heating from the UVB and the diffuse interstellar radiation field, and star formation, while ignoring any form of stellar feedback. These runs allow us to visualize the radial instabilities discussed in Sec.~\ref{Sec:Instabilities}. Importantly, they also serve as a reference for gauging the impact of the NEPS feedback channels implemented in {\tt COLIBRE}. 

As expected, these feedback-free simulations show that the considered gaseous disks are prone to fragmentation, as illustrated in Fig.~\ref{fig:Gas-CoolingSFOnlye}. This figure presents face-on surface density maps of the gaseous disks at $t=500 \rm \ Myr$. At this time, the disks have evolved for several orbital periods, allowing the expected instabilities to develop and form self-bound structures which, in some cases, merge into larger clumps. In the figure, each column shows a different disk mass ($M_{\rm d}$), while panels within a given column show different numerical resolutions, as indicated. The extent of each panel is $20 \rm \ kpc$ across.

The gas disks clearly exhibit a mix of Toomre-unstable and artifically stabilized regions, consistent with the expectations discussed in Section~\ref{Sec:Instabilities}. Moreover, the ability to capture these instabilities depends strongly on resolution. At low resolution, the SPH smoothing kernel artificially smooths density fluctuations, providing a ``numerical'' pressure that suppresses the collapse of structures that would otherwise be physically unstable. This indicates that the disk's stability in these runs is largely governed by numerical, rather than physical, effects.

For the least massive gas disk (leftmost column), the impact of numerical resolution is dramatic. At the coarsest {\tt m6} resolution, the disk appears globally stable, dotted with only a few small clumps. Increasing the resolution to {\tt m5} is sufficient to resolve coherent spiral instabilities, which fragment into the dense clumps that become even more pronounced at the highest {\tt m4} resolution. This demonstrates that while the gas inevitably collapses into dense fragments, the large-scale morphology of the disk is entirely dependent on resolving the underlying physical instabilities.

The intermediate-mass gas disk (middle column) shows a similar trend. At {\tt m6} resolution, fragmentation is apparent, but the large-scale instabilities driving it are poorly resolved. At {\tt m5}, spiral arms broken into dense clumps are clearly visible. However, the inner disk remains artificially stabilized by numerical pressure, an artifact of the finite resolution. This is confirmed in the {\tt m4} simulation, where the instabilities are better resolved, and this artificially stable region shrinks significantly.

In contrast, the most massive disk (rightmost column) is so gravitationally unstable that even the low-resolution {\tt m6} simulation captures its fragmented, clumpy morphology. Here, increasing the resolution to {\tt m5} primarily sharpens the details of filaments and clumps rather than changing the overall structure. Since the disk morphology appears converged, we did not run this case at the numerically expensive {\tt m4} resolution.

Turning to the stellar component, a direct comparison between runs is complicated by the poor numerical convergence of the gas density discussed above. At a fixed disk mass, higher resolution leads to denser gas structures. Since the local star formation rate in {\tt COLIBRE} is linked to the gas free-fall time, these denser disks naturally convert more gas into stars in the absence of feedback from star formation. 

\begin{figure}
    \includegraphics[width=\columnwidth]{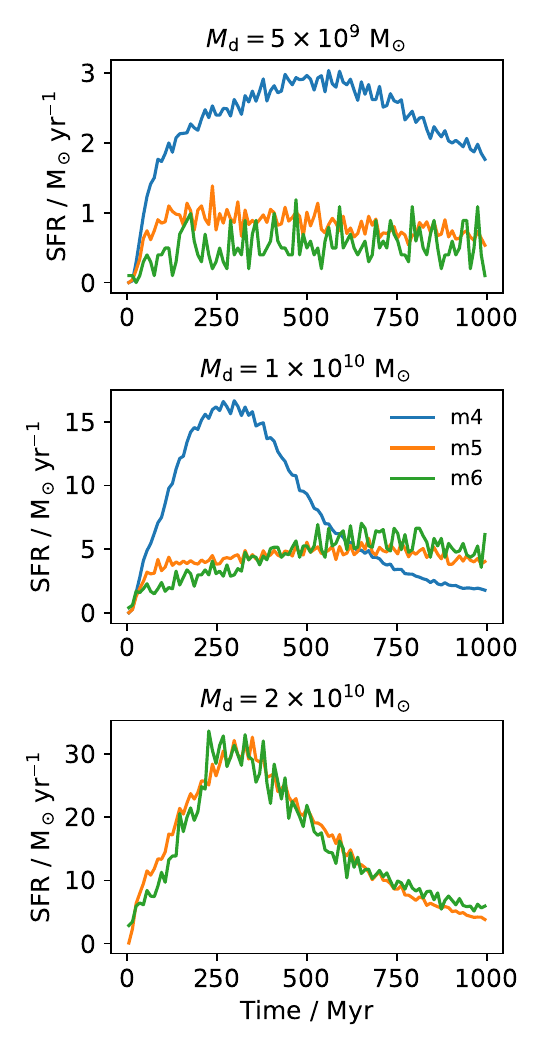}
    \caption{Star formation rate as a function of time, measured in bins equally spaced every 10 $\rm Myr$. From top to bottom, the panels display the results for disks of different gas mass (as indicated), while the different colored lines show results for varying resolutions (as indicated by the legend in the middle panel). These simulations include gas cooling and star formation only, ignoring any source of stellar feedback.}
    \label{Fig:SFR_SF-CoolingOnly}
\end{figure}

\begin{figure*}
    \centering
    \subfloat[Gas density]{
        \includegraphics[width=0.48\textwidth]{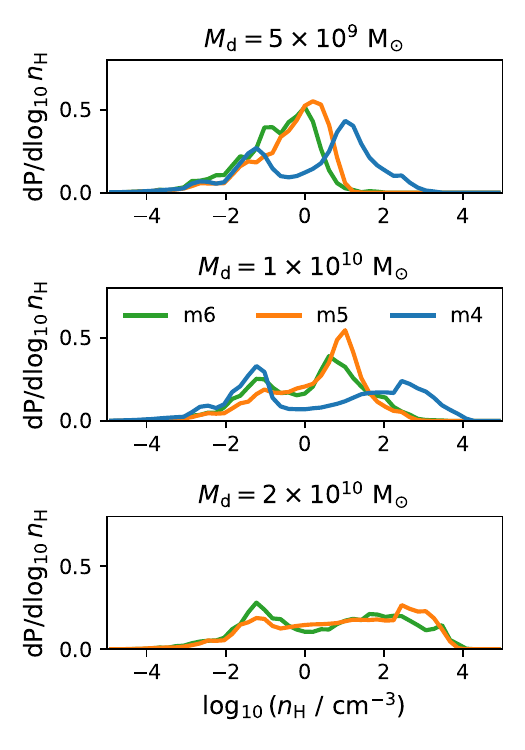}
        \label{fig:rho_hist_woEF}
    }
    \hfill
    \subfloat[Gas temperature]{
        \includegraphics[width=0.48\textwidth]{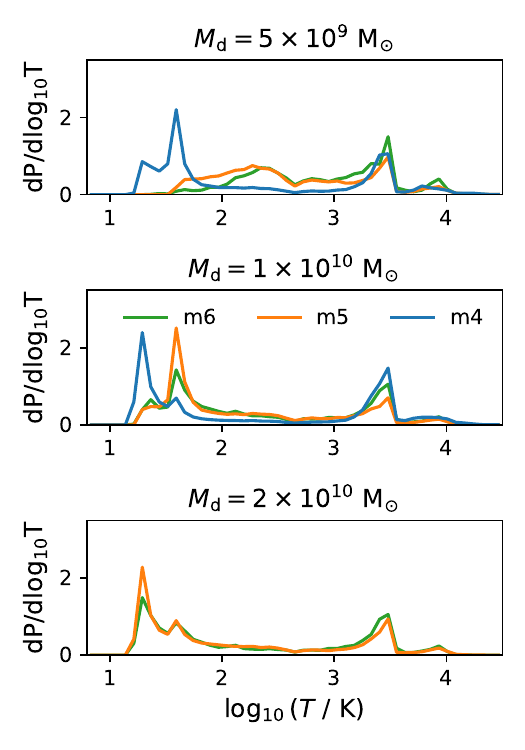}
        \label{fig:T_hist_woEF}
    }
    \caption{Probability distribution functions of the gas density (left) and temperature (right) for the feedback-free simulations at time $t=500$ Myr. These distributions summarize the the thermodynamic state of the gas in the absence of stellar feedback and highlight the profound impact of numerical resolution.}
    \label{fig:rho-T-woEF-hists}
\end{figure*}

Despite these differences in total stellar mass, the morphology of the young stellar disk mirrors that of the gas. This is shown in Fig.~\ref{fig:Stars-CoolingSFOnlye}, which displays stellar surface density maps in the same arrangement as Fig.~\ref{fig:Gas-CoolingSFOnlye}. A comparison reveals that stars trace the gaseous filaments and clumps from which they form. However, the stellar clumps appear systematically larger and more massive than their gaseous counterparts. This is because gaseous clumps are continually formed and depleted by star formation, while stellar clumps persist and grow dynamically through mergers. We have verified that newly formed stars initially trace the gas morphology precisely, but over time, they coalesce into the larger structures seen in Fig.~\ref{fig:Stars-CoolingSFOnlye}, which are distinct from the smaller gaseous clumps still present in the disk.

The star formation histories (SFHs) of these simulations, shown in Fig.~\ref{Fig:SFR_SF-CoolingOnly}, confirm that star formation becomes highly sensitive to numerical resolution for disks artifically stabilized. In the absence of feedback, resolved instabilities lead to runaway gas collapse and an unregulated burst of star formation.

This is particularly evident in the least-massive disk (top panel). At {\tt m6} and {\tt m5} resolutions, where the disk is artificially stabilized, the star formation rate (SFR) is low and steady, below $1 \rm \ M_{\odot} \rm \ yr^{-1}$. At {\tt m4} resolution, however, the simulation resolves the unstable modes, causing the SFR to triple as gas is channeled into dense, rapidly star-forming clumps.

The intermediate-mass disk (middle panel) shows a similar trend. The SFR is modest and nearly resolution-independent at {\tt m5} and {\tt m6}. Once the instabilities are properly resolved at {\tt m4}, the SFR again jumps by more than a factor of three.

In contrast, the most massive disk (bottom panel) is so violently unstable that its SFH converges well across the two explored resolutions. Here, the critical unstable wavelengths are large enough to be resolved even by our coarsest simulations (see Equation~\ref{Eq:lambda_crit} and Fig.~\ref{fig:Radial_instabilities_ICs}). As a result, the SFR is consistently high, peaking at $\approx 30 \rm \ M_{\odot} \rm \ yr^{-1}$, reflecting the rapid and efficient conversion of gas into stars in a massive unstable system. 

The convergence of the SFH in this feedback-free regime may be due to two physical mechanisms. One possibility is that gravitationally-induced turbulence becomes the primary regulator; as the gas disk violently fragments, the strong gravitational interactions between massive clumps stir the gas, creating turbulent motions that act as an effective pressure support against further collapse. Alternatively, the convergence may be a consequence of rapid gas consumption. In this scenario, star formation is so efficient that it is limited by the global availability of fuel, not the resolution of small-scale structures. The disk converts its gas into stars at a maximum rate limited by the identical gas supply in both simulations, naturally leading to a converged SFH. The fact that the SFR peaks early ($\approx 250$ Myr) and then steadily declines supports this gas consumption scenario, though a combination of both mechanisms is plausible.

\begin{figure*}
    \centering
    \subfloat[Face-on gas distribution]{
        \includegraphics[width=0.48\textwidth]{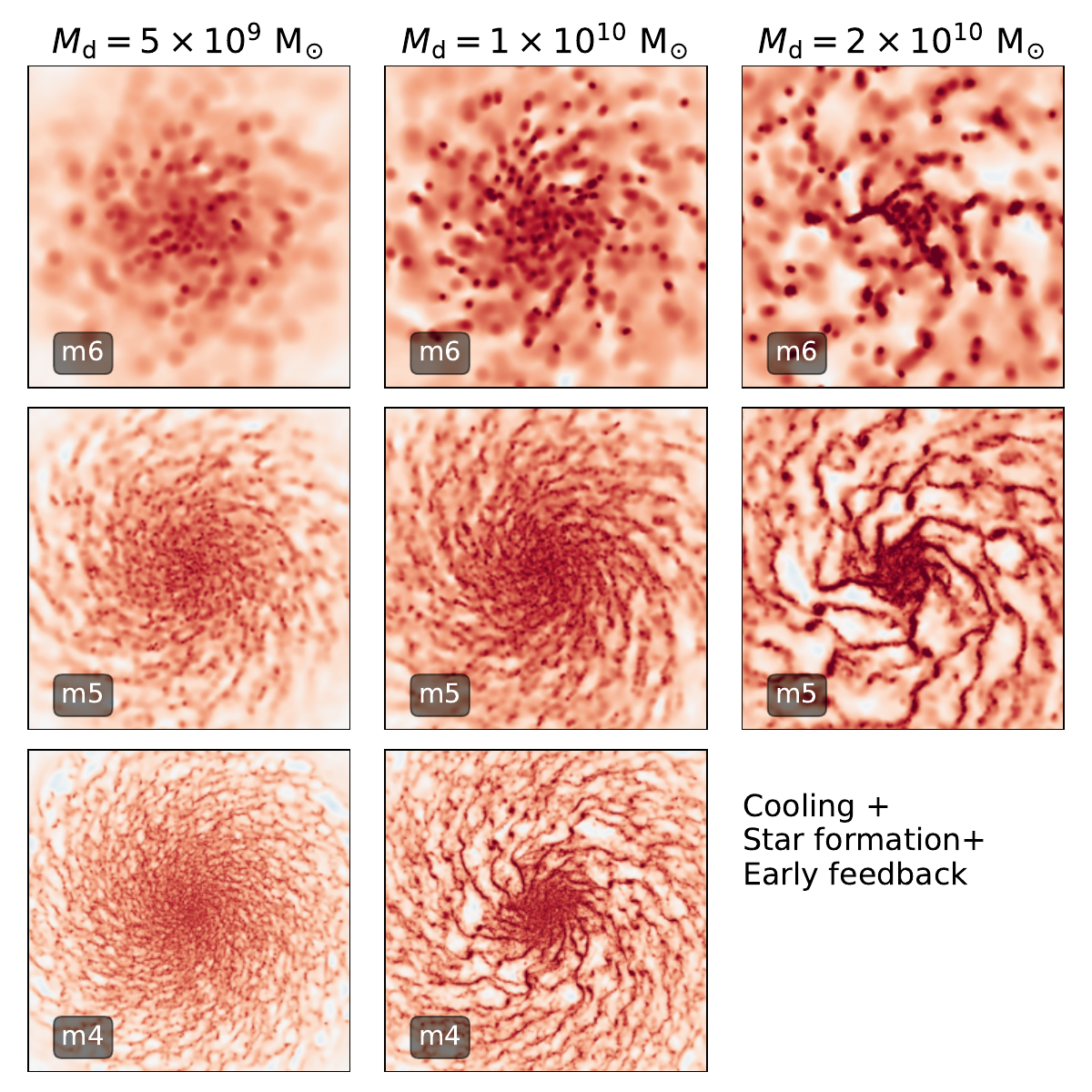}
        \label{fig:Gas-CoolingSF_EarlyFeedback}
    }
    \hfill
    \subfloat[Face-on stellar distribution]{
        \includegraphics[width=0.48\textwidth]{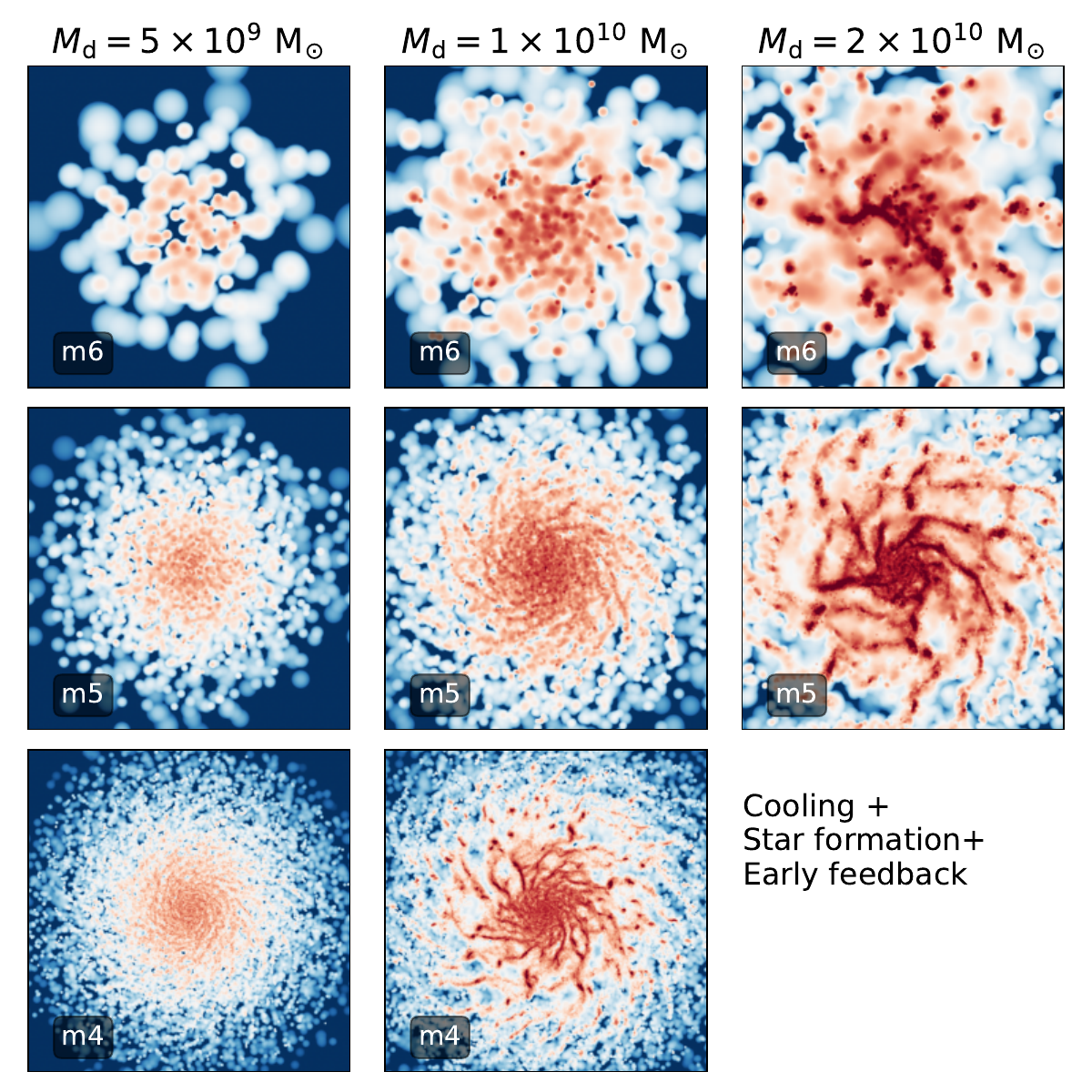}
        \label{fig:Stellar-CoolingSF_EarlyFeedback}
    }
    \caption{Face-on surface density maps of the gas (left) and stellar (right) distribution at $t=500 \rm \ Myr$. The simulations vary the gaseous disk mass, $M_{\rm d}$ (columns), and the gas particle mass resolution (rows). These runs include gas cooling and heating, star formation, and our NEPS feedback implementation. In all cases, supernova feedback is neglected. As in Fig.~\ref{fig:SFOnly-combined}, the color map is idential to that of Fig.~\ref{fig:SFOnly-combined}.}
    \label{fig:CoolingSF_EarlyFeedback-combined}
\end{figure*}

To better understand the physical state of the gas driving these divergent star formation histories, we conclude our analysis of the baseline simulations by examining the gas density and temperature distributions at $t=500$ Myr. These are shown in the probability distribution functions (PDFs) of Fig.~\ref{fig:rho_hist_woEF} (density) and Fig.~\ref{fig:T_hist_woEF} (temperature). In both figures, the top, middle, and bottom panels correspond to the low-, intermediate-, and high-mass gas disks, respectively, while coloured lines indicate the different simulation resolutions.

Fig.~\ref{fig:rho_hist_woEF} illustrates how resolving gravitational instabilities allows gas to collapse to higher densities. For all disk masses, there is a clear trend with resolution: the high-density tail of the PDF extends further as the resolution increases from {\tt m6} to {\tt m4}. This demonstrates that without feedback, higher resolution simulations resolve the collapse of gas into smaller, denser structures. This effect is coupled to the runaway star formation that accompanies it; the rapid consumption of the densest gas in the higher-resolution runs prevents an even larger pile-up of material at the highest densities, but the trend of denser gas being resolved at higher resolution remains robust. This is consistent with the increasingly fragmented and clumpy morphologies seen in Fig..~\ref{fig:Gas-CoolingSFOnlye}.

The temperature PDFs in Fig.~\ref{fig:T_hist_woEF} show the gas separating into a two-phase ISM, with a broad peak at warm temperatures ($\sim 3000$ K) representing diffuse gas and a sharp peak at very low temperatures ($\lesssim 100$ K) for the cold, dense component. For the least massive gas disk, the formation of this cold phase is contingent on resolving the gravitational instabilities that drive gas to high densities. While the amount of cold gas is resolution-dependent, the overall temperature structure is better converged than the density structure. This apparent convergence, however, must be interpreted with caution. The disk's stability and its spectrum of instabilities at this snapshot depend on the total mass of gas remaining, which is in turn sensitive to the disk's prior, resolution-dependent star formation history.

Together, these distributions clarify the outcome of the feedback-free runs: gas attempts to pile up at high densities and low temperatures, but this process is complicated by rapid gas consumption and gravitationally induced turbulence in a manner that is fundamentally dependent on the simulation's resolution and the disk's stability. This runaway behavior reinforces the necessity of physical feedback mechanisms to properly regulate the early stages of star formation in cold galaxy gaseous disks.

\subsection{Gas cooling and heating, star formation, and early feedback}
\label{Sec:early_feedback}

To isolate the impact of our NEPS feedback module, we now analyze a second suite of simulations. These are identical to the baseline runs but now with our early feedback module enabled. SN feedback remains turned off. For these simulations, we set the kick velocity parameter, $\Delta v_{0}=50 \rm \ km \ s^{-1}$, which matches the value adopted for the fiducial {\tt COLIBRE} suite~\citep{Schaye2025}.

The outcome of these experiments is shown in Fig.~\ref{fig:CoolingSF_EarlyFeedback-combined}. In particular, Fig.~\ref{fig:Gas-CoolingSF_EarlyFeedback} displays gas surface density maps directly comparable to those in Fig.~\ref{fig:Gas-CoolingSFOnlye}. A visual comparison immediately reveals that early feedback dramatically alters the disk structure, particularly in simulations that resolve large-scale gravitational instabilities.

\begin{figure}
    \includegraphics[width=\columnwidth]{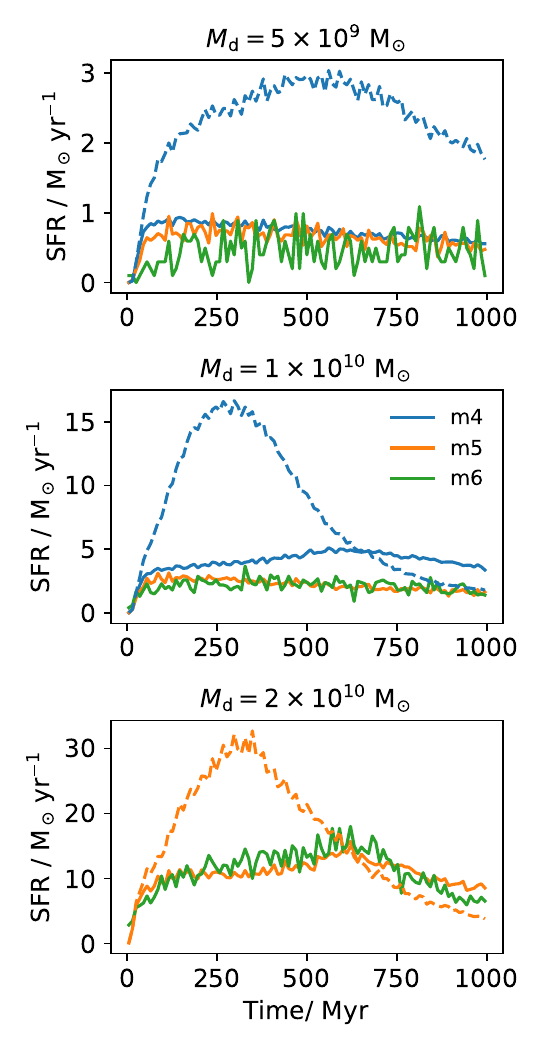}
    \caption{Star formation rate as a function of time, measured in bins equally spaced every 10 $\rm Myr$. From top to bottom, the panels display the results for disks of different gas mass (as indicated), while the different colored lines show results for varying resolutions (as indicated by the legend in the middle panel). These simulations include gas cooling, star formation, and early feedback. For comparison, the dashed lines in each panel show the feedback-free baseline at the highest resolution for each disk, corresponding to the results in Fig.~\ref{Fig:SFR_SF-CoolingOnly}.}
    \label{Fig:SFR_SF-Cooling_EarlyFeedback}
\end{figure}

The effect is best illustrated by the high-resolution ({\tt m4}) intermediate-mass gas disk (middle column). While the feedback-free run shatters into a myriad of small, dense clumps, the run with early feedback sustains large, coherent spiral instabilities. This structural difference stems from two primary self-regulating effects. First, NEPS feedback suppresses star formation, preventing the rapid gas consumption observed in the baseline runs. This preserves a higher overall gas surface density---evident from the redder colors compared to Fig.~\ref{fig:Gas-CoolingSFOnlye}---which helps sustain large-scale instabilities. Second, photoheating from~\ion{H}{II} regions increases the pressure of star-forming regions, stabilizing instabilities against catastrophic collapse into smaller fragments. A similar outcome is observed for the most massive gas disk (rightmost column).

In contrast, where simulations lack the resolution to capture coherent instabilities, the impact of early feedback is minimal. The low-resolution, low-mass gas disk (top-left panel), for instance, appears qualitatively similar with or without NEPS feedback. This is because the disk is already artificially stabilized by numerical pressure, as established in the previous section. With little gas collapsing into dense clumps, star formation is already suppressed numerically, and the physical NEPS feedback mechanisms have little effect on the disk's evolution.

The stellar disks formed in these simulations are shown in Fig.~\ref{fig:Stellar-CoolingSF_EarlyFeedback}. A direct comparison to the feedback-free runs reveals that the stellar distributions are significantly smoother and less fragmented when early feedback is included, although the differences are less dramatic than for the gas disks. The stars also more closely trace the morphology of the gas from which they form. This is because NEPS feedback prevents the runaway collapse and rapid gas consumption seen previously. By regulating star formation, NEPS feedback ensures that the stellar populations are, on average, younger and more closely coupled to the transient gaseous structures, preventing them from merging into the massive, dynamically evolved clumps that dominated the baseline simulations.

To quantify this, we examine the SFHs of these simulations in Fig.~\ref{Fig:SFR_SF-Cooling_EarlyFeedback}. This figure confirms that early feedback successfully suppresses runaway star formation in gravitationally unstable disks. For the high-resolution ({\tt m4}) least-massive gas disk (blue line in the top panel), the excessive star formation seen in the baseline run is significantly reduced. The new simulation now exhibits a steady, low SFR that converges better with its lower-resolution counterpart. This demonstrates that the NEPS feedback module can regulate star formation within the dense, unstable clumps that are resolved at high resolution, even when supernova feedback is disabled.

A similar trend of improved convergence is seen for the intermediate-mass disk (middle panel). However, at the highest resolution, the SFR still steadily increases, indicating that NEPS feedback is not energetic enough to regulate star formation in highly unstable disks. This highlights a key physical limitation of NEPS feedback: while it can effectively suppress star formation within self-gravitating clouds, it lacks the energy to completely disperse larger-scale instabilities. This inability to fully disrupt dense star-forming regions underscores the need for the subsequent, more energetic feedback from core-collapse SNe. 

The interpretation of these results, particularly regarding numerical convergence, must be made with caution due to the unresolved vertical structure of the simulated disks in the absence of efficient feedback. As discussed in Section~\ref{Sec:Instabilities}, the mid-plane volume density, a key input for the {\tt COLIBRE} star formation criterion, is not expected to converge with resolution in our setup, at least when ignoring feedback. Therefore, while the inclusion of NEPS feedback leads to a clear improvement in the convergence of the global star formation rate, this should be understood as the convergence of the full numerical scheme rather than a definitive demonstration of convergence to the true physical solution. However, we note that the additional pressure provided by our NEPS module will likely make the convergence requirements less stringent. A systematic study of these issues is deferred to future work.

\begin{figure*}
    \centering
    \subfloat[Gas density]{
        \includegraphics[width=0.48\textwidth]{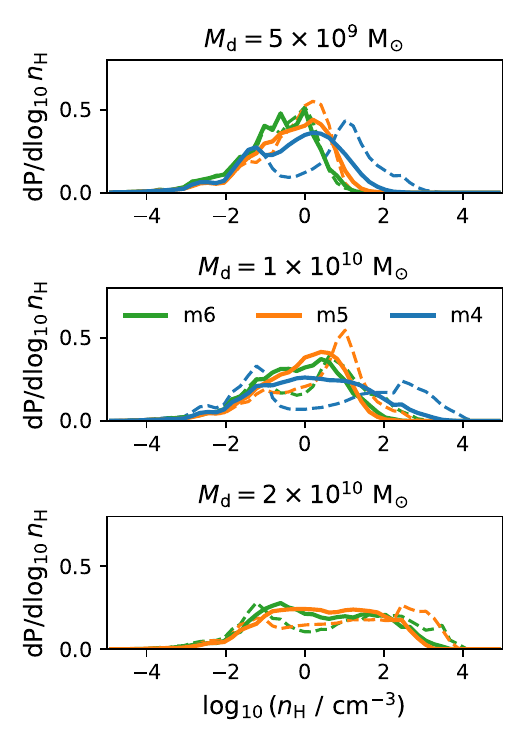}
        \label{fig:rho_hist_wEF}
    }
    \hfill
    \subfloat[Gas temperature]{
        \includegraphics[width=0.48\textwidth]{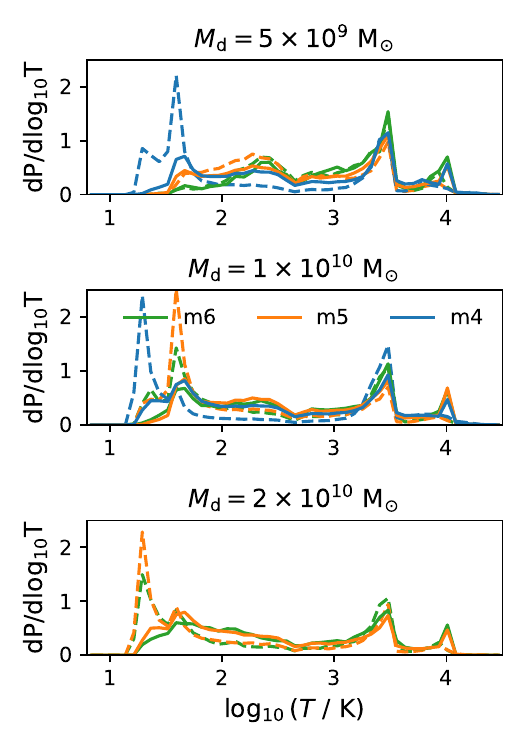}
        \label{fig:T_hist_wEF}
    }
    \caption{As Fig.~\ref{fig:rho-T-woEF-hists} but for simulations that include NEPS feedback (solid lines). These distributions summarize the thermodynamic state of the gas in the absence of SN feedback but including our NEPS feedback. This figures highlights how NEPS feedback dramatically improves the convergence of the simulations. For comparison, the dashed lines show the results for the feedback-free simulations.}
    \label{fig:rho-T-wEF-hists}
\end{figure*}

Finally, we examine the density and temperature PDFs for the simulations including NEPS feedback (Fig.~\ref{fig:rho_hist_wEF} and~\ref{fig:T_hist_wEF}), which are the analogues of those shown for the feedback-free runs in Fig.~\ref{fig:rho-T-woEF-hists}. These figures demonstrate that by successfully regulating star formation, our early feedback module leads to a dramatic improvement in the numerical convergence of the gas thermodynamic state. Because gravitational instabilities can be halted by early feedback, the instantaneous density and temperature structure of the disks is much less sensitive to resolution than in the feedback-free simulations.

However, the limits of this regulation are visible in the intermediate-mass gas disk, where the highest-resolution simulation displays a tail of denser gas not present in the lower-resolution runs. This is a manifestation of the inability of NEPS feedback to fully suppress the onset of gravitational collapse in the densest clumps, signaling the development of large instabilities that our feedback cannot fully prevent. Therefore, the excellent convergence observed for the most massive gas disk is only apparent; it largely stems from the disk's initially violent instability, which makes its structure insensitive to resolution, as is the case for the feedback-free simulations. The fact that star formation is not truly regulated in this case is confirmed by the steadily increasing star formation rate previously observed for this particular example. However, we stress that although star formation is not fully regulated, NEPS feedback reduces the rate at which stars form by more than a factor of three compared to feedback-free simulations.

The gas temperature distribution in the simulations including NEPS feedback converges extremely well with resolution, showing an even more robust result now that early feedback is enabled. Inspection of Fig.~\ref{fig:T_hist_wEF} reveals that the ISM in these simulations has developed a more complex, three-phase structure than was seen in the baseline runs. The gas now populates three distinct thermal phases: a cold phase ($T<100$ K), a warm phase ($T \approx 3 \times 10^3$ K), and a photoionized phase characterized by gas at $T=10^4$ K. This latter temperature is set ``by-hand'' in our subgrid prescription that mimics \ion{H}{II} regions (see Sec.~\ref{Sec:HIIregions-implementation}). In reality, the equilibrium temperature of photoionized gas is not fixed but depends on a balance of heating and cooling processes that are sensitive to the local gas density, metallicity, and dust content. Our choice of a single temperature is a simplification for numerical efficiency. While this effectively models the injection of thermal pressure, a more physically realistic implementation would calculate the temperature self-consistently, likely resulting in a broader distribution of temperatures for the ionized phase. 

\begin{figure}
    \includegraphics[width=\columnwidth]{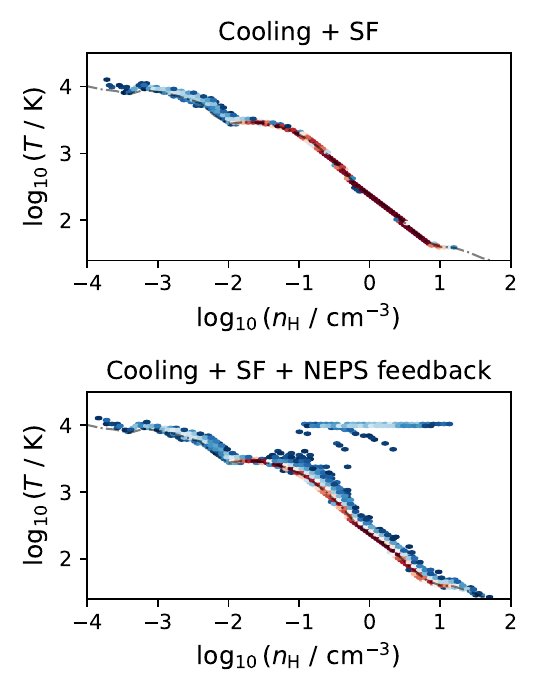}
    \caption{The density-temperature diagram of the gas in the least-massive gas disk at intermediate ({\tt m5}) resolution at time $t=500 \rm \ Myr$. The top panel shows the feedback-free baseline simulation, while the bottom panel incorporates the effects of NEPS feedback (supernova feedback is ignored in both). The primary impact of the early feedback module is the introduction of a warmer, ionized phase to the ISM, characterized by gas at a temperature of $T=10^4$ K set by the subgrid implementation of~\ion{H}{II} regions. The faded dot-dashed line depicts the equilibrium curve for solar metallicity gas. Colours encode the number of gas particles within each hexbin, increasing from blue to red.}
    \label{Fig:rho-T}
\end{figure}

To better illustrate the impact of these~\ion{H}{II} regions, Fig.~\ref{Fig:rho-T} directly compares the density-temperature relation of the gas in the baseline simulation (top) with the run including NEPS feedback (bottom). We specifically show the result for the intermediate-resolution ({\tt m5}) version of the least-massive disk, a case that shows good convergence in both setups. This choice ensures that the differences are a direct consequence of the early feedback physics, not a result of different instabilities developing across the simulations.

The most striking change, readily visible in the comparison, is the emergence of an ionized gas phase at $T=10^{4}$ K, a temperature deliberately set by our model to represent gas within subgrid~\ion{H}{II} regions. This feedback-driven heating creates a clear multiphase structure at densities where star formation occurs. In our model, gas particles are stochastically heated to $10^4$ K, representing their inclusion in a subgrid ~\ion{H}{II} region. Because these particles are subject to physical cooling 2 Myr after they are put in an~\ion{H}{II} region, those in dense environments will cool rapidly. The population of gas particles seen at $10^4$ K therefore represents a dynamic equilibrium, where the rate of  stochastic heating from young stars every 2 Myr is balanced by the physical cooling and recombination rate of the gas. This continuous injection of thermal energy provides a sustained pressure support that regulates star formation and prevents the runaway collapse seen in the feedback-free simulations. The ``plume'' of particles seen between the ionized phase and the equilibrium track represents gas that is in the process of cooling back to equilibrium after leaving their~\ion{H}{II} regions.

Interestingly, although these \ion{H}{II} regions are crucial for regulation, they constitute a small fraction of the total gas mass, as already demonstrated by Fig.~\ref{fig:T_hist_wEF}. 

\begin{figure}
    \includegraphics[width=\columnwidth]{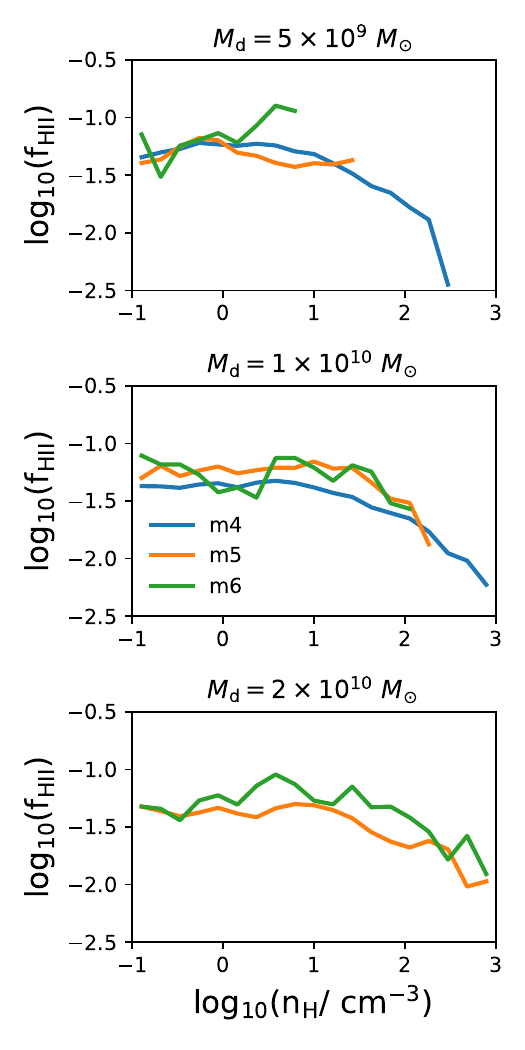}
    \caption{Mass fraction of gas in ~\ion{H}{II} regions as a function of gas density, shown for different gas disk masses (different panels) and resolutions (different coloured lines). The fraction of gas in~\ion{H}{II} regions decreases with increasing gas density. This trend is a direct consequence of the higher recombination rates in denser gas.}
    \label{fig:HIIregions_fraction}
\end{figure}

To quantify the mass contribution of the ionized phase, Fig.~\ref{fig:HIIregions_fraction} shows the instantaneous mass fraction of gas in \ion{H}{II} regions, defined as particles with a temperature $T>10^{3.9}$ K. This figure confirms that although \ion{H}{II} regions are crucial for the regulation of star formation, they are not mass-dominant. They contribute less than $5-10$ \% of the gas mass at any given density, and their fractional importance decreases at higher densities.\footnote{These fractions are expected to decrease even further if the star formation rate decreases, which we expect to be the case if SN feedback is included. While the precise mass fraction in~\ion{H}{II} regions is influenced by our choice of $\Delta t_{\rm HII}$, the qualitative result---that star formation is regulated by a phase that occupies a small fraction of the total gas mass---is physically consistent with the small filling factor of compact~\ion{H}{II} regions in dense molecular clouds~\citep[e.g.][]{Kennicutt1984}}.

This decline is a direct consequence of the higher recombination rates in denser gas, which is not balanced by the higher SFR leading to a higher rate of~\ion{H}{II} regions creation. As shown in Equation~\ref{Eq:HII-mass}, the total mass of the \ion{H}{II} region that can be sustained by a given ionizing flux is inversely proportional to the gas density. Our stochastic implementation captures this physical scaling, leading to a lower probability of gas particles being in the ionized phase at high densities. Consequently, while our early feedback model is critical for regulating star formation, it affects only to a small fraction of the mass. 

\begin{figure*}
    \includegraphics[width=\textwidth]{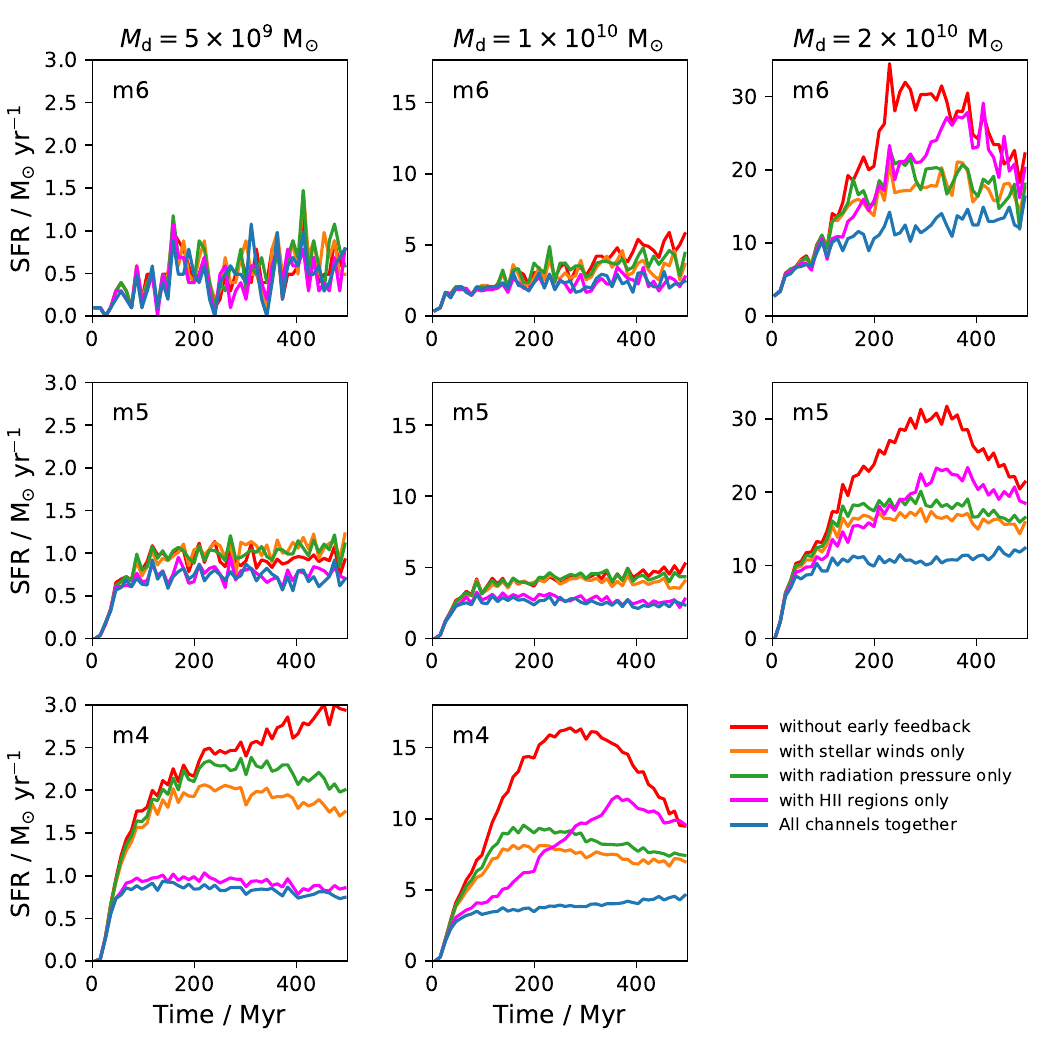}
    \caption{Star formation rate as a function of time, dissecting the impact of the individual components of the NEPS feedback model. The layout is identical to previous figures: columns show simulations with increasing initial gaseous disk mass, $M_{\rm d}$, and rows show increasing numerical resolution from top to bottom. Each panel compares five NEPS configurations: a baseline run with only cooling and star formation (without early feedback; red); runs with only stellar winds (orange), only radiation pressure (green), or only ~\ion{H}{II} regions (magenta); and a run with the full, combined model (with early feedback; blue).}
    \label{fig:SFR_models}
\end{figure*}

To dissect the relative importance of the individual physical processes included in our NEPS feedback model, we run a series of simulations where each component is enabled in isolation. The results are presented in the Fig.~\ref{fig:SFR_models}, which shows the star formation rate as a function of time for simulations with varying disk gas masses (different columns) and numerical resolutions (different rows). Each panel compares the feedback-free case (red) against runs including only stellar winds (orange), only radiation pressure (green), only~\ion{H}{II} regions (magenta), and the full, combined NEPS model (blue). The primary result is immediately clear: in gravitationally unstable disks, the full NEPS model is highly effective at suppressing the runaway star formation that characterizes the feedback-free runs, consistently producing the lowest SFR.

The figure reveals a distinct and consistent hierarchy in the efficacy of the individual feedback channels. The most striking result is the powerful regulatory effect of the~\ion{H}{II} region component. In every unstable disk, the simulation with only~\ion{H}{II} regions active (magenta line) shows a reduction in the SFR compared to the feedback-free case. This highlights the critical role of injecting thermal pressure into star-forming gas, which directly increases the local Jeans mass and provides support against catastrophic gravitational collapse. As discussed in Sec.~\ref{Sec:HIIregions-implementation}, our implementation achieves this via a delayed cooling scheme, which ensures that the thermal energy translates into sustained pressure over a 2 Myr timescale. The dominance of this channel in our model therefore underscores the general importance of sustained pressure support as a primary mechanism for regulating star formation in dense unstable regions.

The kinetic feedback components are considerably less effective on their own. Stellar winds (orange line) provide a moderate level of regulation, while radiation pressure (green line) has the most modest impact, only slightly reducing the SFR. This latter result is consistent with its inefficient coupling with the medium at moderate densities. 

\begin{figure*}
    \centering
    \subfloat[Face-on gas distribution]{
        \includegraphics[width=0.48\textwidth]{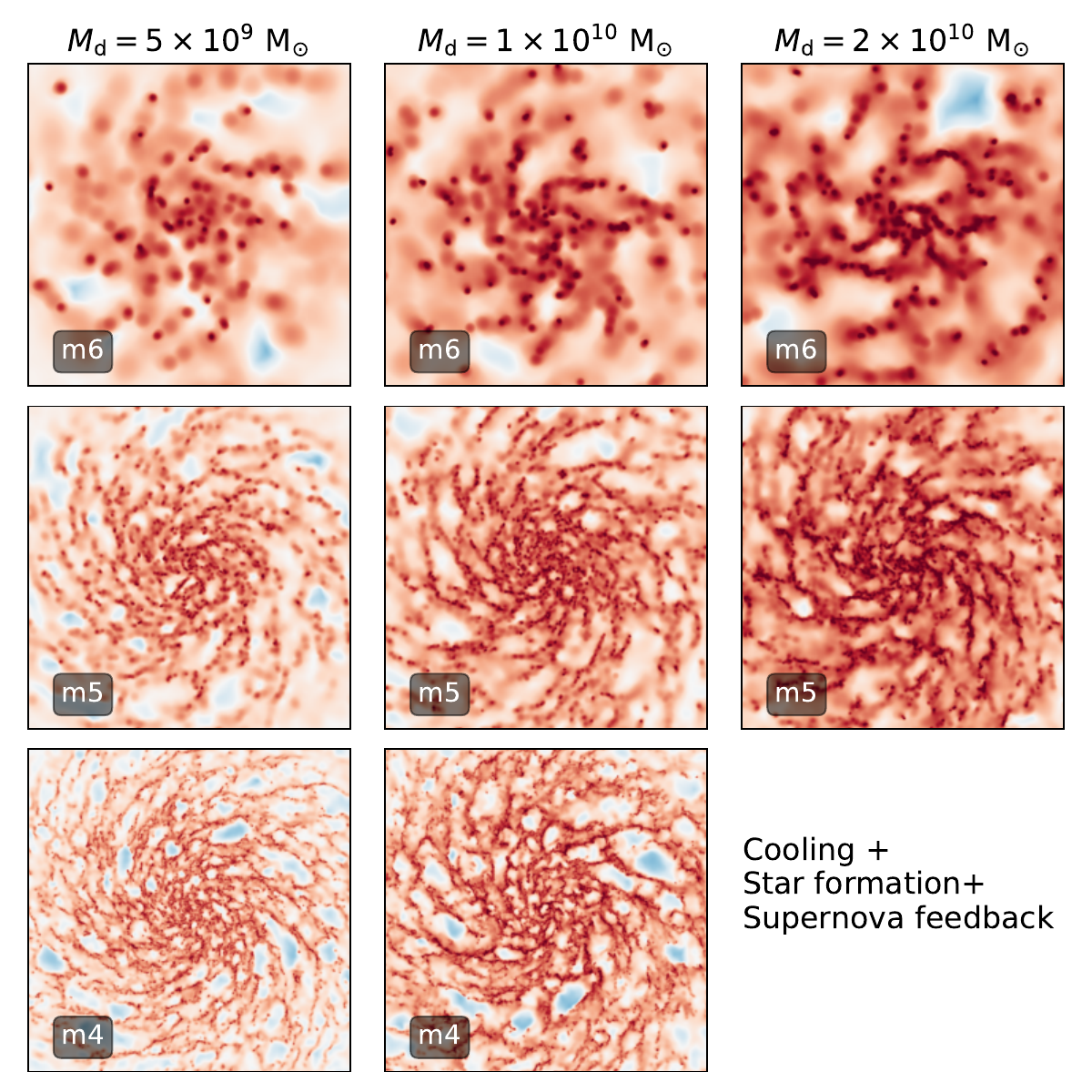}
        \label{fig:Gas-CoolingSF_SNFeedback}
    }
    \hfill
    \subfloat[Face-on stellar distribution]{
        \includegraphics[width=0.48\textwidth]{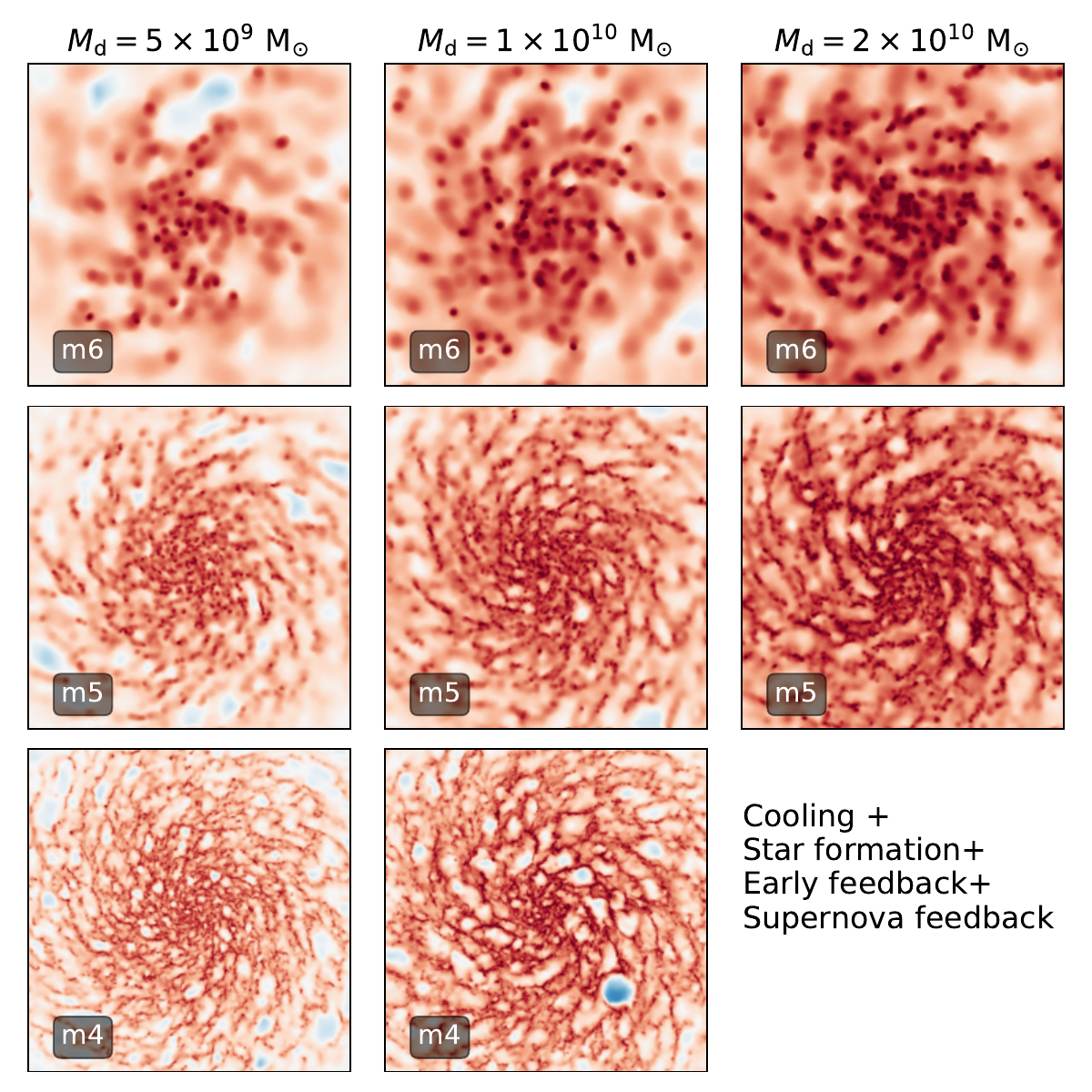}
        \label{fig:Gas-CoolingSF_EarlyFeedback_SNFeedback}
    }
    \caption{Face-on surface density maps of the gas distribution at $t=500 \rm \ Myr$. The simulations vary the gaseous disk mass, $M_{\rm d}$ (columns), and the gas particle mass resolution (rows). The left panels (a) show simulations that include cooling, star formation, and supernova feedback. The right panels (b) show simulations that include cooling, star formation, our NEPS feedback module, and supernova feedback. A comparison reveals that NEPS feedback moderates the impact of SNe, leading to a more homogeneous ISM, particularly in high-resolution simulations. The color map is identical to that of Fig.~\ref{fig:SFOnly-combined}.}
    \label{fig:CoolingSF_EarlyFeedback-combined_SNFeedback}
\end{figure*}

The results demonstrate a powerful synergy between the different feedback mechanisms. The full NEPS model (blue line) consistently achieves a greater level of star formation suppression than even the dominant~\ion{H}{II} region component acting alone. This indicates that the combination of thermal pressure support from photoionization and kinetic momentum injection from winds and radiation provides a more robust and complete form of regulation. This synergy is particularly important for the sustained regulation of star formation at later times, where the full model prevents the gradual rise in the SFR seen in some of the individual feedback runs. The physical origin of this synergy likely stems from the way the kinetic and thermal components interact: momentum injection from stellar winds and radiation pressure drives turbulence in the ISM, preventing gas from collapsing to the highest densities. This shifts star formation to a lower-density environment where, in turn, thermal feedback from~\ion{H}{II} regions becomes more effective, as the ionized regions are physically larger and the gas cooling times are longer. In turn, photoheating from~\ion{H}{II} regions reduces the gas density, increasing the momentum coupling efficiency from stellar winds.

The importance of this regulation also depends on the physical state of the disk. In the low-mass, low-resolution case (top-left panel), where the disk is artificially stabilized, all feedback models have a negligible effect. However, as the disk mass and numerical resolution increase, the simulations better resolve the underlying gravitational instabilities, and the regulatory power of the NEPS model becomes essential for preventing a starburst.

We end by noting that while thermal feedback from~\ion{H}{II} regions is the single most dominant mechanism to prevent initial runaway collapse of the disk, its efficacy can diminish at later times ($t > 300$ Myr) in high-resolution simulations (e.g., the {\tt m4} run of the $10^{10}\,{\rm M}_\odot$ disk). In these cases, without momentum feedback, the gas can eventually settle into high-density configurations where recombination rates are high, reducing the duty cycle of thermal pressurization. However, the hierarchy between processes remains clear. Simulations lacking thermal feedback consistently undergo catastrophic fragmentation and starbursts, whereas those lacking momentum feedback remain largely regulated, albeit with higher variations in their star formation histories. Thus, we identify the thermal channel as the primary regulator that enables the system to reach a stable state, while the momentum channels help regulate that state and prevent the gas secular drift toward higher densities.

\subsection{The inclusion of supernova feedback}
\label{Sec:pre-SN+Feedback}

We end our exploration by turning to the interplay between the relatively gentle early feedback from young stars and the violent, delayed feedback from SNe. To isolate the effects, we contrast two sets of simulations: one employing only supernova (SN) feedback and another that includes both our NEPS module and the subsequent SN events. Fig.~\ref{fig:CoolingSF_EarlyFeedback-combined_SNFeedback}, which mirrors the layout of previous figures, presents the gas disk morphology at $t=500$ Myr for these two scenarios, revealing differences in how the disks evolve.

\begin{figure}
    \includegraphics[width=\columnwidth]{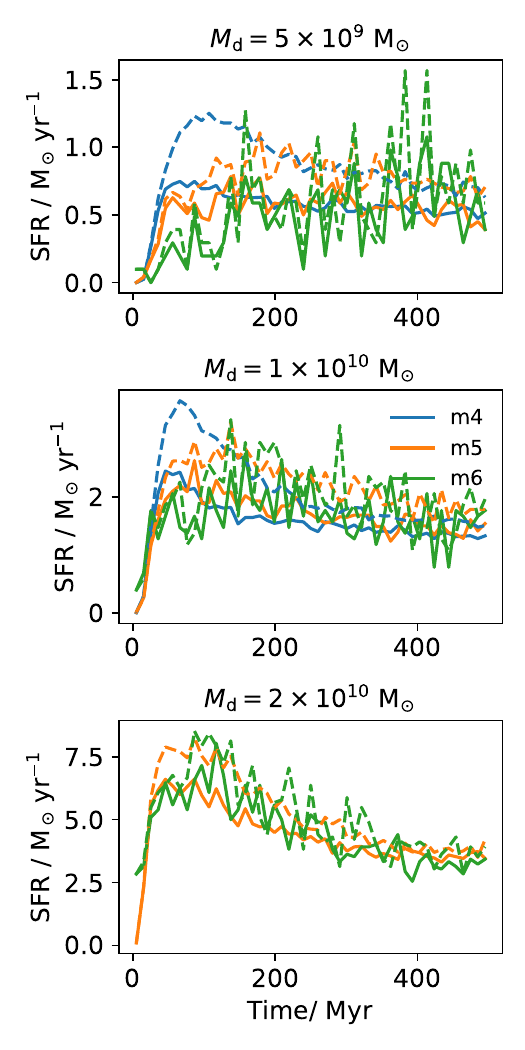}
    \caption{Star formation rate as a function of time. From top to bottom, the panels display results for disks of different gas mass, while different colored lines show results for varying resolutions. Dashed lines correspond to simulations with only supernova feedback. Solid lines correspond to simulations that include both NEPS and supernova feedback. The inclusion of NEPS feedback helps to moderate the initial starburst seen in high-resolution simulations and provides additional regulation on top of supernova feedback.}
    \label{Fig:SFR_SF-Cooling_EarlyFeedback_SN}
\end{figure}

In simulations lacking NEPS feedback, the disk is characterized by large, low-density cavities, seen mostly in blue colours. These structures are the result of highly clustered, energetic SN events occurring in the dense gas near the disk midplane. For instance, the high-resolution intermediate-mass gas disk shows several prominent ``bubbles'' that disrupt the outer disk. In contrast, when the NEPS module is active, the disk is visibly more homogeneous. SN-driven bubbles are systematically smaller and less numerous, indicating that the morphological disruption by SN feedback has been mitigated. However, the global regulation of star formation is enhanced, as shown by the lower SFR in Fig.~\ref{Fig:SFR_SF-Cooling_EarlyFeedback_SN}.

\begin{figure}
    \includegraphics[width=\columnwidth]{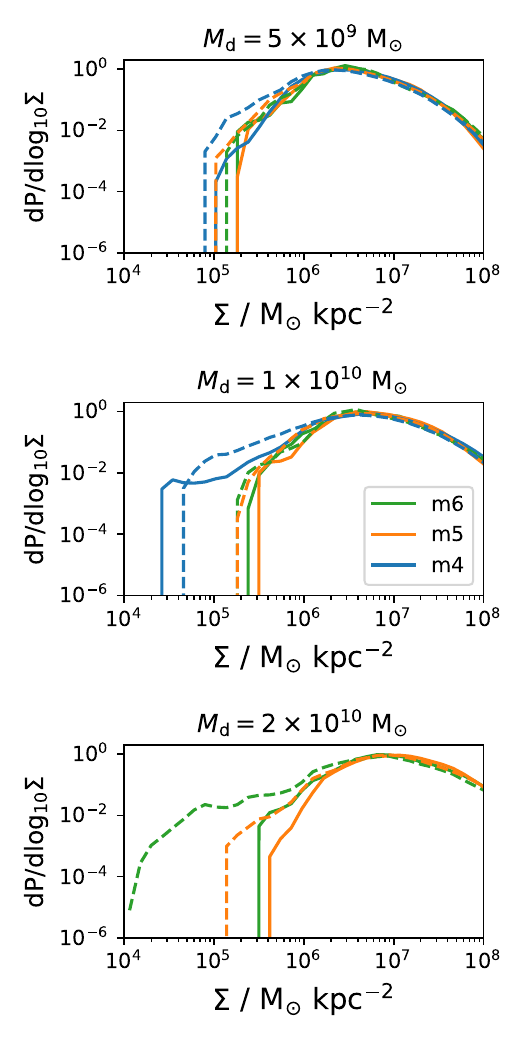}
    \caption{Probability distribution function of the gas surface density, comparing simulations with only supernova feedback (dashed lines) to those that also include NEPS feedback (solid lines) at time $t=500 \rm \ Myr$. Different panels show different gas disk masses, while the colours indicate the numerical resolution. The more prominent low-density tail in the SN-only runs, especially at high resolution, shows that NEPS feedback moderates the impact of SNe, leading to a more homogeneous ISM.}
    \label{fig:surface_density_distribution}
\end{figure}

This qualitative assessment is quantified in Fig.~\ref{fig:surface_density_distribution}, which shows the PDF of the gas surface density, measured directly from the images. For the low-mass gas disks at {\tt m6} and {\tt m5} resolution, the PDFs are nearly identical, confirming that NEPS feedback has little impact on these numerically stabilized systems under the inclusion of SN feedback.

However, a clear difference emerges at {\tt m4} resolution, where the simulation without NEPS feedback (dashed line) exhibits a more prominent tail toward low surface densities. This indicates that the violent, clustered supernova events in that run are more effective at evacuating gas and creating low-density regions. A similar, more pronounced trend is visible for the intermediate- and high-mass disks, confirming that NEPS feedback creates a more homogeneous ISM by moderating the impact of SNe.

The differences across simulations with and without early feedback are a direct consequence of the regulatory effect of NEPS feedback on star formation. By increasing the pressure support within star-forming clouds, early feedback prevents the formation of overly dense stellar clusters where star formation would otherwise proceed with high efficiency. This leads to a less clustered, more spatially, and temporally distributed population of young stars. As a result, subsequent supernova explosions are less correlated, preventing the highly concentrated energy injection required to create large-scale bubbles and stir the disk.\footnote{While this effect is qualitatively evident in the gas morphology and low-density bubbles distribution, a rigorous quantification would require, for instance, calculating the two-point correlation function of supernova events or young stars.}

\begin{figure}
\includegraphics[width=\columnwidth]{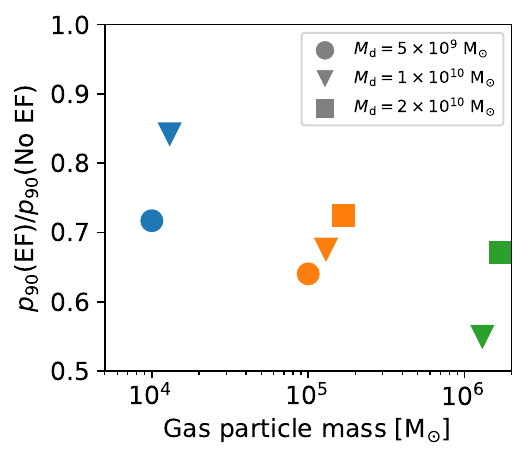}
    \caption{Ratio of the $90$-th percentile of the stellar birth density distribution of simulations with NEPS feedback to those without, where both types of simulations include SN feedback. The horizontal axis shows the numerical resolution, while different symbols correspond to the disk mass, as indicated by the legend. The ratio is consistently below unity, quantitatively demonstrating that early feedback suppresses the formation of stars in extremely dense environments and/or early feedback supresses the formation of very dense gas clouds.}
    \label{fig:Percentiles_ratios}
\end{figure}

Fig.~\ref{fig:Percentiles_ratios} provides quantitative support for this interpretation by showing the ratio of the $90$-th percentile of the stellar birth density distribution between simulations with and without NEPS feedback. Regardless of disk gas mass (symbols) or resolution (x-axis), the ratio is consistently less than one. This demonstrates that NEPS feedback effectively lowers the maximum densities at which stars form, confirming its regulatory role in preventing runaway collapse within star-forming clouds. Among the reasons responsible for this regulation, it is possible that the NEPS feedback suppresses star formation as it increases the turbulence in star-forming regions. This enhanced turbulent pressure would naturally suppress star formation within the {\tt COLIBRE} criterion for star formation (see Sec.~\ref{Sec:Code}).

This picture is further supported by the SFHs presented in Fig.~\ref{Fig:SFR_SF-Cooling_EarlyFeedback_SN}, which compares simulations with only supernova feedback (dashed lines) to those including both NEPS and SN feedback (solid lines). The figure reveals two key results. First, the inclusion of SN feedback, regardless of whether early feedback is active, significantly suppresses the overall SFR, particularly in massive, unstable disks, compared to feedback-free simulations (compare with Fig.~\ref{Fig:SFR_SF-CoolingOnly}), and also compared with simulations that include only NEPS feedback (compare with Fig.~\ref{Fig:SFR_SF-Cooling_EarlyFeedback}). Second, adding NEPS feedback on top of supernova feedback provides further regulation.

This interplay is most evident in the high-resolution (m4) simulations of the low- and intermediate-mass gas disks (blue lines, top and middle panels). Without NEPS feedback, the disk undergoes an initial runaway collapse, producing a prominent starburst that peaks within the first 10 Myr before declining. By contrast, the simulation that includes early feedback shows only a minor initial burst before settling into a steady, well-regulated state that is converged with the lower-resolution runs. For the most massive disk, NEPS feedback similarly helps to moderate the initial starburst and maintains a lower overall SFH, though the effect is less pronounced.

Finally, although not shown here, a qualitative assessment of the morphology of the stellar disks reveals that for the most massive, high-resolution runs, the stellar disk is slightly smoother when NEPS feedback is active. This is a direct consequence of the reduced star formation and higher ISM pressure in these simulations, which prevents the formation of massive, gravitationally dominant stellar clumps.

\section{Discussion}
\label{Sec:Discussions}
As discussed previously, the primary motivation for our NEPS feedback module is to model early physical processes that begin to regulate star formation before the onset of SN feedback, and thus improve numerical convergence of simulations of galaxy formation, particularly at high resolution see Table~\ref{tab:convergence_summary} for a summary of convergence for various physical quantities across the simulated models). As demonstrated in our feedback-free experiments (Figs.~\ref{fig:SFOnly-combined},~\ref{Fig:SFR_SF-CoolingOnly}, and~\ref{fig:rho-T-woEF-hists}), simulations that resolve the cold, dense phase of the ISM are prone to runaway gravitational fragmentation and resolution-dependent star formation histories. Our results show that the implementation of a physically motivated non-explosive pre-supernova feedback module in the {\tt COLIBRE} model provides an effective solution to this problem, enabling more robust and self-regulated galaxy growth over time.

\begin{table*}
	\centering
	\caption{Summary of convergence for key physical quantities across the simulated models.}
	\label{tab:convergence_summary}
	\begin{tabular}{l l l p{7.5cm}} 
		\hline
		\textbf{Physical Quantity} & \textbf{Model Configuration} & \textbf{Status} & \textbf{Physical/Numerical Explanation} \\
		\hline
		Mid-plane Gas Density & All Models & Not Converged & The vertical scale height of the cold gaseous disk remains unresolved even at our highest resolution ({\tt m4}). The maximum density is driven by numerics rather than physics. \\
		\hline
		Star Formation Histories & Feedback-Free & Poor & Simulations undergo runaway gravitational collapse. Higher resolution resolves smaller, denser clumps, leading to notably higher SFRs. \\
		& NEPS (Early Feedback) & Good & Thermal pressure from H\,\textsc{ii} regions regulates the collapse of dense gas. The SFR becomes largely independent of numerical resolution once regulated. \\
		& Full Model (NEPS + SN) & Excellent & The synergy between early regulation and SN feedback prevents both initial starbursts and late-time high SFRs, leading to a highly stable, converged SFH. \\
		\hline
		Gas Temperature Structure & Feedback-Free & Poor & The mass fraction of cold gas ($<100$ K) depends strongly on resolution, driven by resolution-dependent runaway collapse. \\
		& NEPS / Full Model & Good & The inclusion of photoheating establishes a stable three-phase ISM (Cold, Warm, Ionized). The PDF of gas temperatures becomes robust to changes in resolution. \\
		\hline
		Disk Morphology & Feedback-Free & Poor & Low-resolution runs are artificially stabilized by numerical pressure, while high-resolution runs fragment into small dense clumps. \\
		& Full Model (NEPS + SN) & Good & NEPS feedback prevents the formation of overly dense clusters, leading to distributed SNe that maintain a smooth, homogeneous disk structure across resolutions. \\
		\hline
	\end{tabular}
    \label{Tab:Table1}
\end{table*}

Our analysis of the individual components of the NEPS model reveals a clear hierarchy of effectiveness. Thermal feedback from~\ion{H}{II} regions is unambiguously the dominant regulatory mechanism, being responsible for the bulk of the star formation suppression (Fig.~\ref{fig:SFR_models}). By heating a small faction of the gas to 10$^4$ K, this process provides the pressure support needed to counteract gravitational collapse in the dense clumps where stars form. Kinetic feedback from stellar winds and radiation pressure plays a more modest role. While stellar winds provide a secondary level of regulation, the impact of radiation pressure is modest in our model. This latter result must be interpreted in the context of our model's single-scattering approximation, which provides a conservative lower limit on the momentum budget. In the dense, optically thick environments where star formation is most vigorous, the contribution from multiple-scattered infrared photons could boost the efficacy of radiation pressure, a key avenue for future model refinement.

Despite the dominance of~\ion{H}{II} regions, our results also highlight a powerful synergy when all three feedback channels are combined. The full NEPS model consistently produces lower star formation rates than any single component acting alone. The physical origin of this synergy likely lies in the interplay between kinetic and thermal feedback: momentum injection from winds and radiation drives turbulence, preventing gas from collapsing to the highest densities. This shifts star formation to a lower-density regime where thermal feedback from~\ion{H}{II} regions is more effective, as the ionized regions become physically larger and cooling times are longer. Conversely, photoheating from ~\ion{H}{II} regions increases the local pressure, shifting the sites of star formation toward low-density gas, where the momentum injected by stellar winds is more effective at generating turbulence and opposing gravitational collapse. This cooperative process provides a more robust and sustained form of regulation than thermal pressure alone.

We found that NEPS feedback moderates subsequent, more energetic SN events. In simulations with only SN feedback, star formation proceeds in a highly clustered and violent manner, leading to correlated SN explosions that carve out large, low-density cavities and violently disrupt the disk (Fig.~\ref{fig:CoolingSF_EarlyFeedback-combined_SNFeedback}). When NEPS feedback is included, it ``pre-processes'' the ISM by regulating star formation within dense clouds. This prevents the formation of overly compact stellar clusters and leads to a more spatially and temporally distributed pattern of SN explosions. The result is a more homogeneous ISM and a gentler, more continuous star formation, where the destructive power of SNe is reduced. This finding reframes the role of early feedback: its primary function may not be to drive large-scale galactic outflows on its own, but rather to create the conditions necessary for SN feedback to operate more realistically, regulating the galaxy as a whole rather than simply disrupting it locally. This synergistic relationship is fundamental to achieving a self-regulated equilibrium in simulated galaxies, and may contribute to producing thinner gaseous disks compared with simulations that do not include it.

While our NEPS module represents a significant step forward for the {\tt COLIBRE} model, it is important to acknowledge its limitations, which also point toward clear directions for future development. 

First, our idealized disk simulations do not resolve the vertical scale height of the cold gas. As noted in Sec.~\ref{Sec:Instabilities}, this means that the mid-plane volume density---a key input for the star formation law---is not numerically converged. Therefore, while our model dramatically improves the convergence of the global star formation rate (Fig.~\ref{Fig:SFR_SF-Cooling_EarlyFeedback_SN}), this should be understood as the convergence of the full numerical scheme rather than a definitive demonstration of convergence to the true physical solution. Verifying these results in simulations with sufficient resolution to resolve the vertical gas structure remains a critical future step. We note, however, that the increased thermal and turbulent pressure in simulations including SN feedback will relax the numerical resolution requirements for resolving the vertical scale height.

Second, our model for~\ion{H}{II} regions confines feedback to the SPH kernel of individual star particles. This approach does not capture the overlap of multiple~\ion{H}{II} regions from clustered star formation, which in reality would merge to create a much larger ionized region in high-resolution simulations. Our implementation should therefore be interpreted as a ``subgrid pressurization'' scheme rather than a direct simulation of~\ion{H}{II} region expansion. Capturing the large-scale effects is a challenge that will be important to address in future work. Nevertheless, we note that~\ion{H}{II} regions forming in high-resolution simulations are expected to be smaller than a single gas particle.

Third, our estimate of the optical depth neglects the detailed subgrid density structure of the gas. The presence of unresolved dense clumps can enhance the coupling of radiation pressure, or conversely, create low-density channels that allow radiation to leak. However, given the pronounced dominance of the thermal~\ion{H}{II} region feedback channel we expect that even a moderate increase in the radiation pressure coupling efficiency due to subgrid clumping would not overturn the established hierarchy of feedback mechanisms shown in Fig~\ref{fig:SFR_models}.

Finally, this work has focused on idealized disk galaxies to isolate the impact of the NEPS module. The next logical step is to apply this module within large volume simulations, which is done for the {\tt COLIBRE} suite~\citep{Schaye2025}. However, understanding the interplay of the different NEPS ingredients can be more efficiently explored using less expensive ``zoom-in'' simulations. This would allow to study the impact of NEPS on a wider range of galaxy properties, including the circumgalactic medium, the formation of galactic outflows, and the evolution of galaxies across different environments and cosmic epochs.

\section{Conclusions}
\label{Sec:Conclusions}
We have presented the implementation and testing of a new module for non-explosive pre-supernova (NEPS) feedback designed for the {\tt COLIBRE} model of galaxy formation~\citep{Schaye2025}. This work is motivated by observational evidence and by the long-standing challenge of numerical convergence in high-resolution simulations, where the absence of early regulation by stellar feedback allows gas in the cold ISM to undergo runaway gravitational collapse, potentially leading to excessive and resolution-dependent star formation. Our NEPS module mitigates this by incorporating feedback channels from young, massive stars that operate before the first core-collapse SNe. The model implements three key physical processes: the injection of momentum from stellar winds, momentum from radiation pressure, and thermal energy from photoionization heating in~\ion{H}{II} regions. The age- and metallicity-dependent energy and momentum budgets are derived from {\tt BPASS} stellar population synthesis models. Momentum is injected via a stochastic kinetic scheme to ensure it couples effectively to the gas, while the impact of~\ion{H}{II} regions is modeled by stochastically heating gas to $10^4$ K.

We tested our implementation using a suite of simulations of isolated, gravitationally unstable disk galaxies, spanning a range of masses, and evolved at three different numerical resolutions. Our main findings are as follows:

\begin{itemize}
\item In the absence of stellar feedback, our simulations show that gravitationally unstable disks exhibit runaway fragmentation. The resulting star formation rates (SFRs) are highly sensitive to numerical resolution, as higher resolution resolves smaller, denser, and more rapidly star-forming clumps. This leads to a strong lack of convergence in both the galaxy morphology and its integrated star formation history, reinforcing the need for a physical regulatory mechanism.

\item NEPS feedback alone dramatically improves numerical convergence. By providing sustained pressure support in dense, star-forming gas, our model successfully regulates runaway collapse. This results in smoother gas and stellar disks and global SFRs that are significantly lower and less sensitive to resolution. This early feedback establishes a dynamic, three-phase ISM, where a photoionized component ($T=10^4$ K) created by~\ion{H}{II} regions coexists with the warm and cold gas phases. While energetically crucial for regulation, these~\ion{H}{II} regions are sub-dominant in mass, constituting less than 10\% of the gas at any given density.

\item We find clear hierarchy and synergy among the NEPS feedback mechanisms. Analysis of the individual NEPS components reveals that thermal feedback from~\ion{H}{II} regions is the single most dominant regulatory mechanism. Kinetic feedback from stellar winds provides a moderate level of suppression of star formation, while radiation pressure has the most modest impact. Crucially, the full NEPS model is more effective than any single component, demonstrating a powerful synergy where kinetic feedback drives turbulence, shifting star formation to lower-density environments where thermal feedback from~\ion{H}{II} regions is, in turn, more effective, and vice versa.

\item The limitations of NEPS feedback highlight the need for SNe. The physical limits of early feedback become apparent in the most massive and gravitationally unstable disks. At high resolution, we find that NEPS feedback alone is insufficient to fully stabilize the rate of star formation, as it lacks the energy to completely disperse the largest self-gravitating clouds. This results underscores the necessity of subsequent, more energetic feedback from SNe to regulate star formation.

\item NEPS feedback acts as a crucial moderator for supernova feedback. When coupled with the {\tt COLIBRE} supernova module, our NEPS model ``pre-processes'' the ISM. By regulating star formation within dense clouds, it prevents the formation of overly compact stellar clusters. This leads to a more spatially and temporally distributed pattern of subsequent supernova explosions, taming their violent impact and preventing the disruption of the disk. 
\end{itemize}

Although our model successfully addresses its primary goals, we acknowledge its limitations, which are discussed in detail in Sec.~\ref{Sec:Discussions}. Key among these are the use of a single-scattering approximation for radiation pressure (which makes our estimates of radiation pressure coupling in regions with very high dust surface densities a lower limit), and the confinement of~\ion{H}{II} regions to the SPH kernel of the young stars. These represent important avenues for future model development.

In summary, the {\tt COLIBRE} NEPS module provides a physically motivated and numerically robust framework for modeling the first stages of stellar feedback. It successfully mitigates some of the resolution-dependent issues that plague feedback-free simulations by regulating star formation within dense gas clumps. Crucially, it works in synergy with subsequent SN events, creating the conditions for a more continuous and gentle growth of galaxies.

\section*{Acknowledgements}
We acknowledge the anonymous referee for comments and suggestions that helped improve the presentation of our results and further justify our numerical choices. ABL acknowledges support by the Italian Ministry for Universities (MUR) program “Dipartimenti di Eccellenza 2023-2027” within the Centro Bicocca di Cosmologia Quantitativa (BiCoQ), and support by UNIMIB’s Fondo Di Ateneo Quota Competitiva (project 2024-ATEQC-0050). SP acknowledges support by the Austrian Science Fund (FWF) through grant 10.55776/V982. EC and FH acknowledge funding from the
Netherlands Organization for Scientific Research (NWO) through research
programme Athena 184.034.002. JT acknowledges support of a STFC Early Stage Research and Development grant (ST/X004651/1).

\section*{Data Availability}
The data underlying this article will be shared on reasonable request to the corresponding author.

\bibliographystyle{mnras}
\bibliography{references} 


\appendix

\section{Stellar winds kick velocity}
\label{Ap:kick_velocity}

As detailed in Sec.~\ref{Sec:NEPS_model}, our stochastic momentum injection scheme requires defining a target kick velocity, $\Delta v_0$, to determine the probability of kicking neighboring particles. While the total momentum delivered to the ISM is fixed by the {\tt BPASS} model, the choice of $\Delta v_0$ determines how this momentum is sampled. A higher velocity implies kicking fewer particles (i.e., lower mass loading) to conserve the total budget. Here, we explore the sensitivity of our results to this parameter.

\begin{figure}
    \includegraphics[width=\columnwidth]{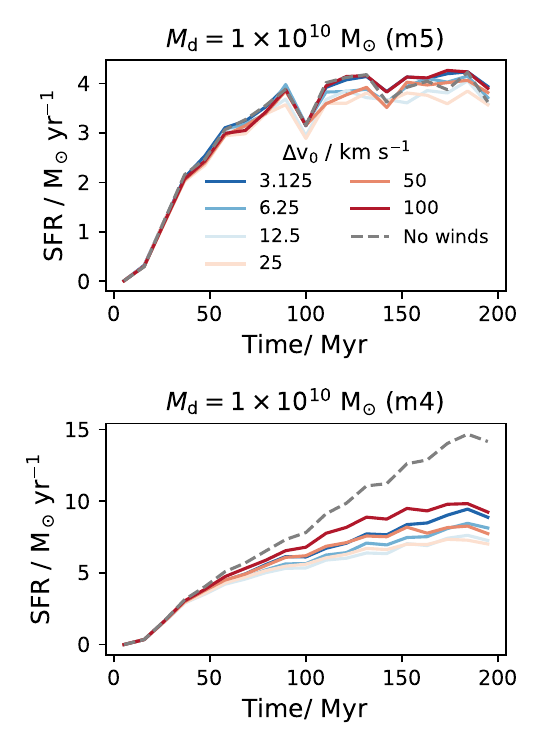}
    \caption{Star formation rate as a function of time for the intermediate-mass disk ($M_{\rm d} = 10^{10} \  \text{M}_\odot$), varying the stellar winds kick velocity parameter, $\Delta v_0$, as indicated in the legend. The total momentum is kept constant in all runs, meaning higher velocities result in fewer particles kicked. The dashed grey line indicates the baseline run without stellar winds. The top panel shows results for the intermediate ({\tt m5}) resolution, for which the SFR is insensitive to the kick velocity because the disk structure is numerically stabilized. In the bottom panel, we show the high-resolution ({\tt m4}) counterpart. Here the disk is both physically and numerically unstable. The SFR is maximally suppressed for wind speeds of $\approx 25$ km s$^{-1}$. Above and below this threshold, the SFR is less supressed, but the differences are small.}
    \label{fig:velocity_kick}
\end{figure}

Figure~\ref{fig:velocity_kick} shows the star formation history of the $10^{10}\,\text{M}_\odot$ disk for $\Delta v_0$ ranging from $\approx 3$ to $100$ km s$^{-1}$. The top panel shows results at our lower ({\tt m5}) resolution. In this regime, the disk is artificially stabilized as discussed in Sec.~\ref{Sec:Instabilities}. Consequently, the momentum from stellar winds is expected to have a negligible impact, which is indeed what this panel shows; the SFR is only weakly dependent on the chosen kick velocity. Simulations spanning nearly two orders of magnitude in $\Delta v_0$ produce comparable star formation histories. 

However, a very subtle, non-monotonic trend is visible. Increasing the kick velocity up to 25 km s$^{-1}$ slightly decreases the average SFR at later times, as expected. Yet, for values exceeding 25 km s$^{-1}$, the SFR begins to rise again, becoming comparable to runs without stellar winds (shown by the grey dashed line). This occurs because the probability of kicking any given gas particle is inversely proportional to $\Delta v_0$; for extremely high kick velocities, the momentum is imparted to so few particles that its overall regulatory effect on the disk diminishes.

The bottom panel of Figure~\ref{fig:velocity_kick} shows a similar test at the higher ({\tt m4}) resolution, where the disk is numerically unstable and early feedback is essential for regulation. For all values of $\Delta v_0$ explored, we confirm that stellar winds have a significant impact on the SFR, suppressing it by a factor of $\approx 2$ at later times. Despite this, a clear trend with $\Delta v_0$ emerges. Similar to the trend observed for the {\tt m5} counterpart, velocity kicks between $\approx 12$ to $25$ km s$^{-1}$ maximize the suppression of the SFR. For kick velocities $\lesssim 12$ km s$^{-1}$ (blue/cyan lines), the winds become increasingly ineffective. This is likely because the velocity imparted to the star-forming clouds is comparable to or less than the local effective sound speed, rendering the kicks dynamically less relevant. 

For higher velocities, $\gtrsim 50$ km s$^{-1}$, the star formation rate increases with the kick velocity. As with the lower resolution case, this occurs because conserving the total momentum with a higher target kick velocity requires a decrease in mass loading. At very high velocities, the scheme kicks only a few particles very hard; these high-speed ejecta may escape the local dense region without efficiently coupling their momentum to the bulk of the cloud.

Despite this second-order non-monotonic behaviour, the variations in SFR between models with different kick velocities are relatively modest. Given that thermal pressure from~\ion{H}{II} regions remains the dominant stabilizing mechanism, our fiducial choice of $\Delta v_{0}=50$ km s$^{-1}$ places our model in the robust, effective regime. 

\section{\ion{H}{II} regions lifetime}
\label{Ap:HII_time}

\begin{figure}
    \includegraphics[width=\columnwidth]{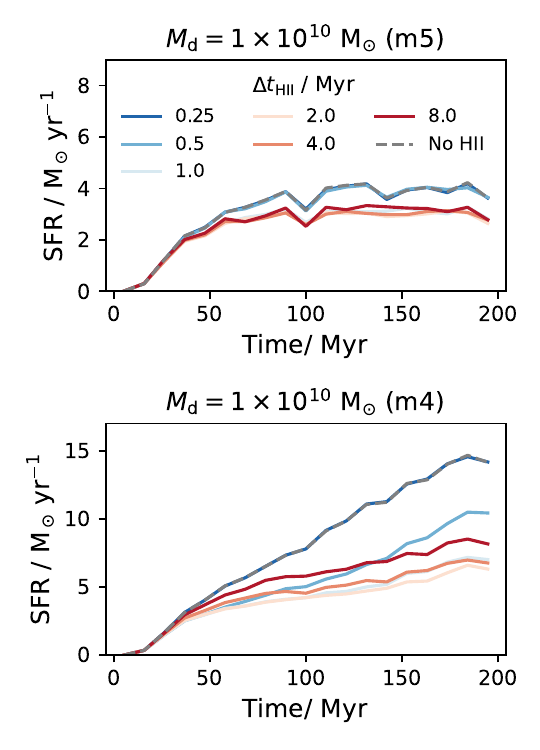}
    \caption{Star formation rate as a function of time for the intermediate-mass galaxy ($M_{\rm d} = 10^{10} \ \text{M}_\odot$), varying the~\ion{H}{II} region lifetime parameter, $\Delta t_{\ion{H}{II}}$. The dashed grey line indicates the baseline run without \ion{H}{II} regions. The top panel shows the intermediate-resolution ({\tt m5}) simulations, where the disk is stabilized by numerical pressure and results are insensitive to the parameter choice. The bottom panel shows the high-resolution ({\tt m4}) counterpart. For timescales shorter than the sound-crossing time across the resolution element, the thermal energy suffers from numerical overcooling and fails to regulate star formation. Robust regulation is achieved only when $\Delta t_{\ion{H}{II}}$ is long enought to allow the thermal feedback to produce mechanical P dV work to expand the locally unstable star-forming clouds. For the {\tt m4} resolution, we find that $\Delta t_{\ion{H}{II}} = 2$ Myr is a good compromise.}
    \label{fig:HII_time}
\end{figure}

In Sec.~\ref{Sec:HIIregions-implementation}, we argued that a fixed timescale of $\Delta t_{\ion{H}{II}} = 2$ Myr is required to avoid numerical overcooling in our simulations. To validate this, we performed a parameter exploration using the $M_{\rm d} = 10^{10} \ \text{M}_\odot$ disk. The top panel of Fig.~\ref{fig:HII_time} shows the results at intermediate ({\tt m5}) resolution. As discussed in previous sections, the disk is numerically stabilized in this regime, rendering the star formation history largely insensitive to the details of the feedback implementation.

The bottom panel of Fig.~\ref{fig:HII_time} shows the results at the higher ({\tt m4}) resolution, where the disk is numerically unstable and regulation is essential. Here, varying $\Delta t_{\ion{H}{II}}$ from $0.25$ to $8.0$ Myr reveals a sharp transition in feedback efficiency. For $\Delta t_{\ion{H}{II}} \le 1.0$ Myr, the SFRs gradually approach the ``No~\ion{H}{II} feedback'' baseline (dashed line), indicating that these simulations are dynamically equivalent to those without~\ion{H}{II} feedback. Indeed, the run with $\Delta t_{\ion{H}{II}}= 0.25$ Myr is virtually identical to that without early feedback. This is a clear manifestation of the numerical overcooling problem, where the injected thermal energy is radiated away before it can expand the gas. In contrast, for $\Delta t_{\ion{H}{II}} \ge 1.0$ Myr, the feedback successfully suppresses the SFR by a factor of $\approx 2$ compared to the feedback-free run, and the results converge with $\Delta t_{\ion{H}{II}}$. This confirms that the heating duration must be comparable to the sound-crossing time across the SPH kernel of the dense star-forming clouds to ensure physical coupling.

While $\Delta t_{\ion{H}{II}}$ is fixed at 2 Myr in this work and in the {\tt COLIBRE} simulation suite, we anticipate that as numerical resolution approaches the scales of individual molecular clouds, this parameter should depend on density and eventually converge toward the physical recombination timescale. However, for the resolutions explored in our work, which are typical of large-volume simulations, $2$ Myr represents a physically motivated timescale that prevents artificial energy loss while stabilizing unstable disks.

\bsp	
\label{lastpage}
\end{document}